\documentclass[aip,jcp,reprint,amsmath,amssymb]{revtex4-2}

\usepackage{physics}
\usepackage{textalpha}
\usepackage{xcolor}

\usepackage[version=4]{mhchem} 
\usepackage{graphicx}
\usepackage{tabularx}
\usepackage{booktabs}
\usepackage{multirow}
\usepackage{diagbox}
\usepackage{subcaption}
\usepackage{lineno}
\usepackage[ruled,vlined]{algorithm2e}
\usepackage{hyperref}
\usepackage{listings}
\usepackage{cleveref}
\usepackage{mathtools}
\usepackage{notes2bib}
\bibnotesetup{record-type = footnote}

\def\selectlanguage#1{}

\newcommand*\citeit[1]{\citeauthor{#1}\cite{#1}}
\newcommand*\citeitcomma[1]{\citeauthor{#1},\cite{#1}}
\newcommand*\citeitperiod[1]{\citeauthor{#1}.\cite{#1}}

\begin{document}

\title{Automatic dual basis sets for self-consistent field calculations}

\author{Joonatan Huhtasalo}
\affiliation{Department of Chemistry, University of Helsinki, P.O. Box
  55 (A. I. Virtasen aukio 1), FI-00014 Helsinki, Finland.}

\author{Susi Lehtola}
\email{susi.lehtola@alumni.helsinki.fi}
\affiliation{Department of Chemistry, University of Helsinki, P.O. Box
  55 (A. I. Virtasen aukio 1), FI-00014 Helsinki, Finland.}

\date{\today}

\begin{abstract}
  Density functional theory (DFT) has long been the workhorse of electronic structure theory, and it is experiencing a new boom thanks to big data applications and machine learning.
  A DFT calculation requires the iterative solution of the self-consistent field (SCF) problem, whose cost is the product of (i) the number of SCF steps and (ii) the cost of a single step.
  The number of SCF steps depends strongly on the initial guess, regardless of the SCF algorithm, and quantitative DFT accuracy is known to require at least a polarized triple-$\zeta$ basis set.
  However, a qualitatively correct solution may often be found already in a smaller \emph{dual basis} (DB), which can then bootstrap a calculation in the full basis set.
  Traditionally, the DB is chosen by hand for each application to enable robust convergence to the ground state.
  In this work, we pursue automatic generation of DBs for SCF calculations.
  Our generator constructs a pruned orbital basis from the target orbital basis for a predefined precision threshold and a given Fock matrix.
  As expected, the larger the basis set is, the smaller the fraction of kept basis functions.
  The DBs produced by our algorithm exhibit similar performance for a wide range of systems from the first 4 periods of the periodic table.
  Ultimately, we find that presently available initial guesses for the Fock matrix are not accurate enough to unlock the use of adaptive basis sets, and believe that the same limitation may hold for similar attempts for basis compactification.
\end{abstract}

\maketitle


\section{Introduction}
\label{sec:intro}

Density functional theory\cite{Hohenberg1964_PR_864, Kohn1965_PR_1133} (DFT) has become the workhorse of computational chemistry and materials science, offering an attractive balance of accuracy and cost.
Its reliability and affordability have made it the engine of high-throughput screening campaigns and the data source for training novel machine-learning models.
Both of these types of applications routinely demand millions of electronic structure calculations.\cite{Curtarolo2013_NM_191, Butler2018_N_547}

In this big-data regime, the speed and robustness of self-consistent field (SCF) calculations become essential, since each DFT calculation means solving an SCF problem.
The time for an individual SCF calculation depends on (i) $N_{\rm{iter}}$, the number of cycles required to reach convergence, and (ii) $t_{\rm{cycle}}$, the time for one SCF cycle, regardless of the employed SCF algorithm.
Even modest savings per calculation may thus accumulate into large savings of computer time and electricity in big-data applications.

A number of strategies in the literature have unlocked big data applications of DFT and continue to make calculations more affordable in the future.
The first strategy is to make each step cheap to evaluate.
As the cost of SCF is traditionally integrals-bound, much effort has been spent on optimizing the cost of Fock matrix builds over the years through various approaches, such as improved integral algorithms,\cite{Dupuis1976_JCP_111, McMurchie1978_JCP_218, Obara1986_JCP_3963, HeadGordon1988_JCP_5777, Gill1994_AQC_141, Reine2012_WIRCMS_290} the use of density fitting\cite{Whitten1973_JCP_4496, Baerends1973_CP_41, Dunlap1979_JCP_3396, Vahtras1993_CPL_514, Eichkorn1995_CPL_283, Weigend2002_PCCP_4285} and later on Cholesky decompositions,\cite{Beebe1977_IJQC_683, Koch2003_JCP_9481, Pedersen2024_WIRCMS_1692} numerical quadrature,\cite{Becke1988_JCP_2547, Murray1993_MP_997, Treutler1995_JCP_346, Stratmann1996_CPL_213} seminumerical approaches,\cite{Friesner1985_CPL_39, Neese2009_CP_98, Plessow2012_JCC_810, Laqua2018_JCTC_3451} and general purpose graphics processing units (GPGPUs).\cite{Yasuda2008_JCC_334, Ufimtsev2008_JCTC_222, Ufimtsev2009_JCTC_1004, Asadchev2010_JCTC_696, Kalinowski2017_JCTC_3160, Barca2020_JCTC_7232, Seritan2020_WIRCMS_1494}

Another, complementary strategy is to make the SCF iteration itself converge more reliably through robust solvers:\cite{Francisco2004_JCP_10863, Francisco2006_JMC_349, Slattery2024_PCCP_6557, Sethio2024_JPCA_2472, Fdez.Galvan2025_JCTC_12674, Lehtola2025_JPCA_5651, Greiner2026_JCTC_881} an extra investment of time per iteration by an improved algorithm can pay itself off by a large reduction in the number of necessary iterations, especially in the case of systems with complicated electronic structure.
However, even the most robustly converging algorithm can only be expected to locate the minimum closest to the starting point, which need not be the global minimum; obtaining the correct solution with such an algorithm thus hinges on the availability of a good initial guess, which can be also supplemented by metadynamics,\cite{Thom2008_PRL_193001} for example.

The third strategy is to develop better and more automated initial guesses.
We argue that this is an increasingly important route to making large-scale DFT both faster and more robust.
When provided with a guess that is already close to a minimum, SCF algorithms reach convergence in far fewer and cheaper steps.\cite{Almloef1982_JCC_385, Vacek1999_CPL_189, VanLenthe2006_JCC_32, Shoji2014_CPL_50, Lehtola2019_JCTC_1593, Lehtola2020_JCP_144105, Hegely2022_IJQC_26782}
In contrast, a poor starting guess will (i) in the best case require many additional SCF iterations to be taken, (ii) in a bad case fail to converge altogether, or (iii) in the worst case, silently converge onto an incorrect solution, that is, a saddle-point or a (far) higher-lying local minimum.

In our terminology, an initial guess is a way to generate a suitable
starting guess for SCF calculations, without having to perform any SCF
on the polyatomic system. To be even qualitatively correct, an initial
guess must be able to reproduce atomic shell
structure. State-of-the-art initial guesses include the superposition
of atomic densities\cite{Almloef1982_JCC_385, VanLenthe2006_JCC_32}
(SAD) as well as the superposition of atomic
potentials\cite{Lehtola2019_JCTC_1593, Lehtola2020_JCP_144105} (SAP),
which build on atomic calculations, only, and often land on orbitals
and an energy that is not terribly far from the converged ones (the
guess is of correct symmetry, for example).  Machine learning models
have also recently found use as SCF guesses.\cite{Schuett2019_NC_5024,
  Shao2023_NC_6281, Hazra2024_JCTC_4569, Stishenko2026_JCP_64123,
  Zhang2026_DD_1868}

This work is centered not around initial guesses, per se, but around
accelerating SCF calculations with dual basis (DB) approaches.  Since the
time for a single SCF step scales heavily as a function of the size of
the basis set, it is long-standing practice to seed a large-basis SCF
calculation by the SCF solution found in a small, DB set. A
related, and mostly complementary approach is to pursue the SCF
solution by a divide and conquer strategy based on molecular
fragments.\cite{Mezey1995_JMC_141, Yang1995_JCP_5674,
  Inadomi2002_CPL_139, Chen2005_JCP_184105, Szekeres2006_CPL_420,
  Khaliullin2006_JCP_204105, Jansik2009_JCTC_1027,
  Jansik2009_PCCP_5805, He2010_JCTC_405, Zhang2016_IJQC_357,
  Saether2017_JCTC_5282, Hoeyvik2019_MP_1626929, Wang2021_JCTC_4831,
  Goletto2021_JCTC_7416, Marrazzini2021_JCTC_791,
  Ballesteros2022_JCTC_179, Yu2025_JCTC_1230}

The smaller the DB is, the faster the SCF runs in it, but the
quality of the found solution is also poorer. The small DB set
is conventionally chosen by hand for each application. In this work,
we attempt to automate its choice from the employed orbital basis set,
forgoing the need for expertly tailored basis set pairings.

Our algorithm determines a DB as a suitable subset of basis
functions from the full basis set for a given initial guess for the
Fock matrix. The SCF problem is solved in this DB, and the
orbitals are padded to the full basis (no projection is
necessary). The solution thereby obtained can then be used either as
an initial guess in the full basis, or with various DB
methodologies that approximate the relaxation in the full basis, which
will be discussed below.

The layout of this work is as follows.
We start out by summarizing the relevant theoretical background in \cref{sec:theory}.
We describe our adaptive DB (ADB) algorithm and its implementation in \cref{sec:implementation}.
We present the computational details of this work in \cref{sec:compdet}, listing the software we used, and including details on the studied systems and basis sets.
We then present our results in \cref{sec:results}, and conclude with a summary and discussion in \cref{sec:discussion}.
Hartree atomic units are used throughout the text, unless specified otherwise.

\section{Theory}
\label{sec:theory}

The Kohn--Sham energy functional\cite{Kohn1965_PR_1133} depends on the sought-for molecular orbitals (MOs) $\psi_i(r)=\langle {\bf r} | \psi_i \rangle$.
For a computational solution, the unknown MOs are expanded in a known basis set $\chi_\alpha(\vb{r})$ as
\begin{align}
  \label{eq:lcao}
  \psi_{i}(\vb{r})=\sum_{\alpha}C_{\alpha i} \chi_\alpha(\vb{r}),
\end{align}
where $C_{\alpha i}$ is the $\alpha$-th expansion coefficient of the $i$-th MO.
In this work, we use contracted Gaussian basis functions
\begin{align}
  \label{eq:cgto}
  \chi_\alpha(\vb{r}) = \sum_{k} d_{kj} g_{k}(\vb{r}; \alpha_{kj}),
\end{align}
where $d_{kj}$ are the contraction coefficients, and the primitive Gaussian with exponent $\alpha_{kj}$ is given in the spherical form as
\begin{align}
  \label{eq:pgto}
  g_{k}(\vb{r}; \alpha_{kj})=N_{kj}r^{l}e^{-\alpha_{kj}r^2}Y_{lm}(\theta,\phi),
\end{align}
which has been expressed centered at the origin for simplicity.

As is well known,\cite{Lehtola2020_M_1218} varying the Kohn--Sham functional with respect to the orbital coefficients leads to the Roothaan--Hall equations\cite{Roothaan1951_RMP_69, Hall1951_PRSAMPES_541} for the molecular orbital coefficients (a necessary condition for an extremum)
\begin{align}
 \label{eq:rh}
  \vb{F} \vb{C} &= \vb{SCE},
\end{align}
where $\vb{F}$ is the Kohn--Sham--Fock matrix, $\vb{S}$ is the overlap matrix, and $\vb{E}$ is a diagonal matrix of the corresponding orbital energies.
The problem in DFT calculations is that $\vb{F}$ depends on the occupied orbitals through the density matrix
\begin{align}
  \label{eq:densmat}
  P_{\mu\nu} = \sum_{i} n_i C_{\mu i} C_{\nu i}
\end{align}
built from the orbitals $\vb{C}$ and their occupations $\vb{n}$,
$\vb{F} = \vb{F}(\vb{P})$, since the electrons interact with each
other.  This is the reason why the solution of \cref{eq:rh} begins
from an initial guess, and the reason why iterative methods are
required to find the SCF orbitals.  Simply put, the SCF algorithm then
undertakes the iterative solution $\vb{C}_0 \to \vb{F}_1 \to \vb{C}_1
\to \dots$ until convergence.  The initial guess can be given either
as (i) the pair of the orbitals $\vb{C}_0$ and the corresponding
orbital energies $\vb{E}_0$ to produce occupations $\vb{n}$ with the
Aufbau theorem, (ii) a direct guess for the density matrix $\vb{P}$,
or as a guess for the Fock matrix $\vb{F}_0$, since it yields guess
orbitals $\vb{C}_0$ and orbital energies $\vb{E}_0$ with \cref{eq:rh}.

The orbital dependence of the Fock matrix arises from electronic Coulomb and exchange-correlation contributions.
Even though large (polarized triple-$\zeta$ or larger) basis sets can be required for quantitative accuracy in electronic structure calculations, already a smaller basis set may be able to qualitatively describe the form of these contributions and therefore the overall orbital shape.
The established strategy of a read-in orbital guess therefore employs a self-consistent solution of \cref{eq:rh} in a smaller basis, which is then projected to the target basis for use as the initial guess for SCF.

When the small basis affords a solution that is close to that in the full basis, the remaining orbital relaxation in the large basis can further be approximated instead of solving it self-consistently, giving rise to dual-basis methods,\cite{Havriliak1983_JACS_4, Havriliak1984_CJP_1336, JurgensLutovsky1991_CPL_451, Wolinski2003_JCP_9497, Liang2004_JPCA_3206} which have been recently extended also to perturbation theory.\cite{Maschio2020_JCTC_340, Maschio2024_JPCA_10020}
Dual-basis methods can produce computational savings of up to an order of magnitude,\cite{Liang2004_JPCA_3206} but their cost-accuracy tradeoff again fully depends on the pairing between the target basis and the DB.

In this work, we interest ourselves in the case where the guess orbitals are determined in a subbasis of the target basis.
Our ${\bf C}_0$ will thus have many rows that are filled with zeros.
We then want to find a suitable sparsity pattern in the solution to \cref{eq:rh}: to identify its important structures by adaptively finding a compact set of basis functions that leads to eigenvectors and eigenvalues that are close to those of its full solution.

Given that there are $N! / [(N-K)!K!]$ ways to choose a subset of $K$ functions out of a pool of $N$ basis functions, the exact solution to this problem is intractable.
Furthermore, each candidate would ideally have to be ranked in the resulting SCF energy, which would be prohibitively expensive.
Instead of trying to find the best solution, we will be satisfied with one that is simply good enough to be usable.

We therefore investigate the use of the guess Fock matrix ${\bf F}_0$ to choose the DB: if the Fock matrix guess is qualitatively correct, it should be able to distinguish between the energetic importances of the various basis functions in any given system.
The criterion we will optimize for a given ${\bf F}_0$ is the occupied-space sum $E_{\rm occ}$ of the orbital energies $E_i$ 
\begin{align}\label{eq:criterion}
  E_{\rm{occ}}({\rm DB};\vb{F}_0) = \sum_{\sigma\in\{\alpha,\beta\}}\sum_{i}^{N^{\sigma}_{\rm{occ}}}E_{i}({\rm DB};\vb{F}_0),
\end{align}
which is motivated by \cref{eq:rh} and the Aufbau theorem, where the sum runs over the $N^\sigma_{\rm occ}$ occupied orbitals of each spin $\sigma \in \{\alpha, \beta\}$.
The orbital energies $E_{i}({\rm DB};\vb{F}_0)$ are obtained by solving \cref{eq:rh} in the DB for the given $\vb{F}_0$ in the full basis.
We will thus develop an algorithm that chooses the DB by minimizing \cref{eq:criterion}.

In practice, \cref{eq:rh} is solved with L\"owdin's symmetric\cite{Loewdin1950_JCP_365} or canonical orthogonalization\cite{Loewdin1956_AP_1} procedure.
Simply put, one finds an orthogonalizing matrix $\vb{X}$ that satisfies $\vb{XSX}^{\rm T} = 1$, and rewrites the unknown MOs as $\vb{C} = \vb{X \tilde{C}}$, whereby \cref{eq:rh} reduces to a standard eigenvalue problem
\begin{equation}
  \label{eq:rh-ev}
  \tilde{\vb{F}}   \tilde{\vb{C}} = \vb{E} \tilde{\vb{C}}
\end{equation}
where $\vb{\tilde F} = \vb{X}^{\rm T}\vb{F}\vb{X}$.
Now, selecting a subbasis compresses $\vb{\tilde F}$ onto a subspace, $\vb{\tilde F}_\text{sub} = \vb{W}^{\dagger}\vb{\tilde F}\vb{W}$ with $\vb{W}^{\dagger}\vb{W}=\vb{I}$, and the Poincar\'e separation theorem then bounds the compressed eigenvalues $\{E'_{i}\}$ from below by those of the full matrix,
\begin{align}
  \label{eq:cauchy}
  E_{i} \leq E'_{i} \quad \forall i\in\{1,2,\dots,\dim(\vb{\tilde F}_\text{sub})\},
\end{align}
with the eigenvalues sorted in ascending order.
Cauchy interlacing is the special case in which $\vb{W}$ merely selects coordinates, so that $\vb{W}^{\dagger}\vb{\tilde F}\vb{W}$ is a principal submatrix.

The above mathematical argument shows that $E_{\rm occ}(\vb{F}_0)$ is monotonically decreasing, enabling the use of $E^{\rm occ}$ to rank candidate basis functions: one can simply use a greedy algorithm that includes the function(s) that lower $E^{\rm occ}$ the most at every step.
The truncation error in the occupied orbital energy is then given by
\begin{equation}
  \label{eq:deta}
  \Delta E_{\rm occ}(\vb{F}_0) = E_{\rm occ}(\text{DB};\vb{F}_0) - E_{\rm occ}(\text{full basis};\vb{F}_0) \ge 0,
\end{equation}
while the analogous truncation error in the SCF energy is
\begin{equation}
  \label{eq:descf}
  \Delta E_{\rm SCF} = E_{\rm SCF}(\text{DB}) - E_{\rm SCF}(\text{full basis}) \ge 0.
\end{equation}

\section{Implementation}
\label{sec:implementation}

As was already stated above, we will minimize \cref{eq:criterion} with respect to the DB with a greedy algorithm.
However, adaptive improvements only make sense if the occupied orbitals are already qualitatively correct, which is why we decided to initialize the DB with a minimal basis set.
First, we compute the overlap matrix
\begin{equation}
  \label{eq:M}
 M_{Jk} = \int\dd^3 r \chi_{J}^{*}({\bf r}) \chi_{k}({\bf r}) 
\end{equation}
between functions $\chi_J$ in the target basis, and functions $\chi_k$ in the minimal basis set.
$M_{Jk}$ measures the similarity of $ \chi_{J} $ and $ \chi_{k} $.
Looping over all $k$ functions in the minimal basis, we include the function $J$ with maximal $|M_{Jk}|$ in the DB, zeroing out the $J$-th row of $M$ after the inclusion to ensure the same function is not picked again.
After this procedure, the DB holds $K$ functions like the employed minimal basis, chosen for maximal resemblance.

\begin{figure*}
  \includegraphics[width=.8\linewidth]{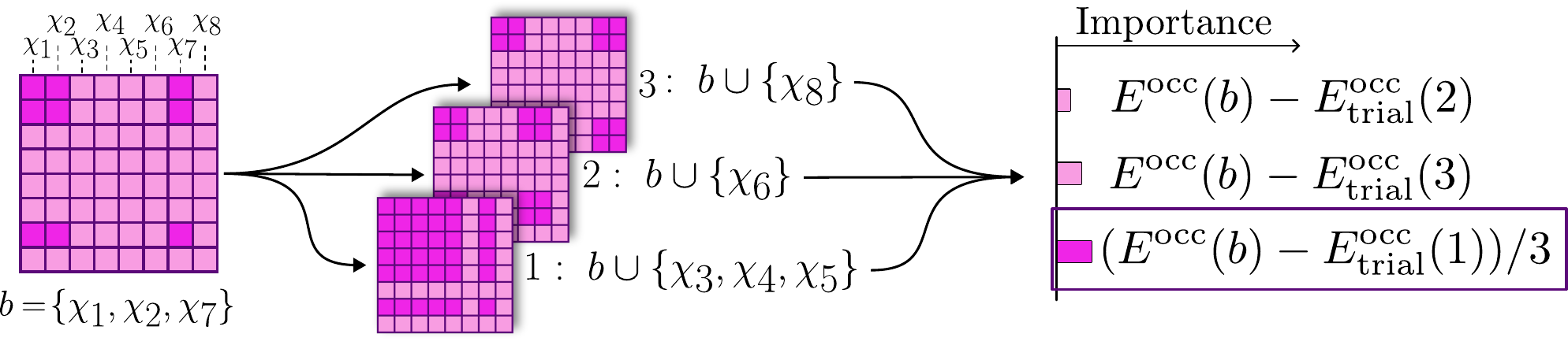}
  \caption{
    Illustration of function selection during one ADB cycle.
    Matrix elements in the darker color correspond to basis functions present in the DB, and are thus a part of $\vb{F}_{0}^{\rm{trial}}$.
    The left panel depicts the starting point: a basis consisting of $\chi_{1}, \chi_{2}$ and $\chi_{7}$.
    The middle panel depicts the formation of trials: (1) the addition of a $p$-shell containing $\chi_{3}, \chi_{4}$, and $\chi_{5}$, (2) the addition of an $s$ shell containing $\chi_6$, or (3) an $s$ shell containing $\chi_8$.
    The sum of occupied orbital energies is computed for each trial, and the decrease is normalized by the number of functions added (3, 1, and 1, respectively).
    The shell that leads to the largest normalized decrease is included in the DB (here, trial 1), and the algorithm proceeds to the next loop iteration.
  }
  \label{fig:abs_cycle}
\end{figure*}

Next, we can proceed with the greedy algorithm presented in \cref{alg:abs} that iteratively solves \cref{eq:rh} in a subbasis, refining the solution.
The algorithm needs the overlap matrix $\vb{S}$ and the guess Fock matrix $\vb{F}_{0}$.
The size of the subbasis increases at every iteration, as more functions are included.
Each macroiteration in the ADB algorithm consists of
\begin{enumerate}
\item looping over all candidate functions: functions of the full basis that have not yet been included in the DB,
\item evaluating the decrease in \cref{eq:criterion} achieved by including the candidate function(s) in the DB (solve \cref{eq:rh}),
\item stopping the algorithm if the value of \cref{eq:criterion} has converged,
\item including the function(s) that led to the most decrease in the DB and going back to step 1.
\end{enumerate}
Steps 1 and 2 are presented pictorially in \cref{fig:abs_cycle}.

\begin{algorithm}
  \caption{A pseudocode implementation of the ADB algorithm.}
  \label{alg:abs}
  \SetKwRepeat{Do}{do}{while}
  \newcommand{\idxset}{\{i_{\ell,n}^{M}\}}
  \newcommand{\trialinds}{\{j_{\ell,n}^{M}\}}
  \newcommand{\trialidx}{\{j^{\rm{trial}}\}}
  \newcommand{\trial}{b^{\rm{trial}}}
  \newcommand{\tcctiny}[1]{\tcc{\footnotesize #1}}

\small
\DontPrintSemicolon
    \tcctiny{Finds an adapted subbasis of the given basis set suitable for solving the problem approximately.\linebreak
    Set theoretic notation is used to refer to the basis sets and their function indices: $B$ refers to the set of function indices in the full basis (present implicitly in the basis projection and complement of $b$), $b$ to the set of function indices in the DB, etc.
    Therefore the algorithm is general for both cases where the functions are considered (i) one-by-one and (ii) grouped by the shell.}
    \vspace{.4em}
    \KwData{$\vb{F}_{0}, E^{\rm occ}_{\rm{tol}}, B$}
    \KwResult{$b$ \tcc*{\footnotesize Index set of selected functions}}
    \Begin{
        \tcctiny{Initialize basis from minimal basis}
        $b \longleftarrow$ Projection\;
        \tcctiny{Initialize criterion $E^{\rm occ}$ of \cref{eq:criterion}}
        $E^{\rm occ} \leftarrow E^{\rm{occ}}(\vb{F}_{0}\texttt{[}b\texttt{]})$\;

        \While{$b \neq B$}
           {\upshape
             \tcctiny{Save old value}
             \texttt{$E^{\rm occ}_{\rm old}$} $\leftarrow E^{\rm occ}$\;
             \tcctiny{Initialize list of trials}
          \texttt{trials} $\leftarrow$ \texttt{[ ]}\;
            \tcctiny{Loop over available shells}
            \ForEach{$s \in (B \setminus b)$}{
              \tcctiny{Construct list of basis functions}
              $\trial \leftarrow \{f \in s \} \cup b$\;
              $N \leftarrow $ number of basis functions added to the trial basis\;
              \tcctiny{Extract $\vb{F}_{0}$ in the trial basis}
              $\vb{F}_{0}^{\rm{trial}} \leftarrow \vb{F}_{0}\texttt{[}\trial\texttt{]}$\;
              \tcctiny{Compute trial criterion}
              $E^{\rm occ}_{\rm{trial}} \leftarrow E_{\rm{occ}}(\vb{F}_{0}^{\rm{trial}})$\;
              \tcctiny{Append trial to list}
              \texttt{trials.append}($\texttt{[}\trial, N, (E^{\rm occ}_{\rm trial} -E^{\rm occ}_{\rm old})/N\texttt{]}$)\;
            }
            \tcctiny{Find trial that leads to largest normalized reduction of energy}
            $\texttt{idx} \leftarrow \texttt{argmin}(\texttt{trials[:][3]})$\;
            \tcctiny{Update the criterion change}
            $\Delta E^{\rm occ}_{\rm iter} \leftarrow \texttt{trials[idx][2]} \times \texttt{trials[idx][3]}$\;
            \tcctiny{If tolerance is achieved, break}
            \If{$\abs{\Delta E^{\rm occ}_{\rm iter}} \leq E^{\rm occ}_{\rm{tol}}$}{
              \texttt{break}\;
            }
            \tcctiny{Update the basis; shell linking also implemented here}
            $b \leftarrow b \cup \left(b^{C}\texttt{[idx]}\right)$\;
            \tcctiny{Update the current estimate of $E^{\rm occ}$ (affected by shell linking)}
            $E^{\rm occ} \leftarrow E^{\rm{occ}}(\vb{F}_{0}\texttt{[}b\texttt{]})$\;
        }
        \Return{$b$}
    }
\end{algorithm}

We implemented the ADB algorithm of \cref{alg:abs} in Python in an open-source Python library, which is freely available on GitHub.\bibnote{J. Huhtasalo, ADBlib, \url{https://github.com/NMQC-from-HEL/adblib}, 2026. Accessed 25 Nov 2026.}
Although repeated solutions of \cref{eq:rh} in incrementally larger subspaces could be implemented efficiently by iterative updates, for the present proof of concept we employ full diagonalization at every step.
    
As we operate on atomic orbital basis sets, basis functions belonging to the same shell are considered together and handled as one unit.
For example, all functions on the $2p$ shell ($2p_x$, $2p_y$, and $2p_z$) of a hydrogen atom are tested simultaneously, and all three are either included or excluded together.
As a shell with angular momentum $\ell$ contains $N = 2 \ell+1$ or $N = (\ell +1)(\ell + 2)/2$ functions in the spherical and Cartesian case, respectively, we normalize the change in $\Delta E^{\rm occ}$ by $N$ in the comparison to prevent the DB from being unnecessarily dominated by high-angular-momentum functions.

As typical applications employ the same orbital basis set for all atoms of the same element, our solver is also able to employ this restriction in the construction of the DB.
If, for instance, the algorithm decides to add a $3d$ shell on the fifth fluorine atom of a system, the same shell is also added to all other fluorines in the system, when this feature is enabled.
This shell linking is employed in all the calculations of this work.

\section{Computational Details}
\label{sec:compdet}

We used PySCF\cite{Sun2020_JCP_24109} version 2.11.0 for all calculations with the Perdew--Burke--Ernzerhof (PBE) functional\cite{Perdew1996_PRL_3865, Perdew1997_PRL_1396} and a level 7 quadrature grid.
Libcint~\cite{Sun2015_JCC_1664} was used within PySCF to compute Gaussian integrals, and Libxc~\cite{Lehtola2018_S_1} to evaluate the density functional.
Python version 3.10.14 was utilized as the language of choice.

Jensen's polarization-consistent basis sets\cite{Jensen2001_JCP_9113, Jensen2002_JCP_7372, Jensen2002_JCP_9234, Jensen2004_JCP_3463} have been optimized for the employed density-functional level of theory, and our calculations employ the double-$\zeta$ to quadruple-$\zeta$ aug-pc-$n$ basis sets ($n\in\{1,2,3\}$).
The STO-3G~\cite{Hehre1969_JCP_2657, Collins1976_JCP_5142} basis set is used as the minimal basis to seed the iterative refinement of the DB.

We assess our DB generator using the molecular database of \citeitcomma{Weigend2005_PCCP_3297} which consists of 311 molecules representing nearly all elements in most of their common oxidation states.
\citeit{Weigend2005_PCCP_3297} optimized their geometries at the BP86/def2-SV(P) level of theory;\cite{Becke1988_PRA_3098, Perdew1986_PRB_8822} we use the geometries as given.
As the systematic Jensen basis sets are limited to the first four periods of the periodic table, our study considers the 186 molecules for which the Jensen sets are available.
Furthermore, ferrocene was excluded from the study due to size limitations, totaling 185 molecular systems.

Since the database of \citeit{Weigend2005_PCCP_3297} does not report spin states, the ground spin states were first determined in the def2-TZVP basis set.\cite{Weigend2005_PCCP_3297}
The PBE energy of each system was calculated for all the spin multiplicities up to the octet, and the state with the lowest energy was assigned as the ground state.
The resulting identification of molecules with non-singlet ground states is given in \cref{tbl:non-singlet-ground-states}; all other molecules are run as closed-shell singlets.

\begin{table}
  \begin{tabular}{ll|ll|ll|ll}
    Molecule & $M$ &    Molecule & $M$ &     Molecule & $M$ &     Molecule & $M$ \\
    \hline
    \hline
    \ce{AlN} & 3 & \ce{FeF3} & 6 & \ce{MnO} & 6 & \ce{S2} & 3 \\
    \ce{CoCl3} & 5 & \ce{FeO} & 5 & \ce{MnO2} & 4 & \ce{ScO} & 2 \\
    \ce{CoF2} & 4 & \ce{GaF2} & 2 & \ce{MnS} & 6 & \ce{SeO} & 3 \\
    \ce{CoF3} & 5 & \ce{GaO} & 2 & \ce{NiF2} & 3 & \ce{TiF3} & 2 \\
    \ce{CrCl3} & 4 & \ce{GeF3} & 2 & \ce{NiF3} & 4 & \ce{TiO} & 3 \\
    \ce{CrF3} & 4 & \ce{MgF} & 2 & \ce{NiO} & 3 & \ce{VO} & 4 \\
    \ce{FeF2} & 5 & \ce{MnF2} & 6 & \ce{NiS} & 3 & & \\
    \hline
  \end{tabular}
  \caption{Spin multiplicities $M$ for molecules not run as singlets in this study.}
  \label{tbl:non-singlet-ground-states}
\end{table}

For the guess Fock matrix $\vb{F}_0$, we consider the following options:
\begin{enumerate}
\item the converged Fock matrix from the full basis SCF (SCF) \label{enu:scf}
\item the superposition of atomic densities\cite{Almloef1982_JCC_385, VanLenthe2006_JCC_32} ('\texttt{atom}' guess in PySCF) (SAD) \label{enu:sad}
\item the \texttt{minao} variant of the SAD approach used as the default guess in PySCF: the atomic occupied orbitals are taken from the cc-pVTZ basis set\cite{Dunning1989_JCP_1007} (MINAO) \label{enu:sadminao}
\item the extended H\"uckel guess\cite{Hoffmann1963_JCP_1397} following the implementation of \citeitcomma{Lehtola2019_JCTC_1593} which employs atomic orbitals and orbital energies as proposed by \citeitcomma{Norman2012_CPL_229} and either the generalized Wolfsberg--Helmholz formula\cite{Wolfsberg1952_JCP_837} to estimate off-diagonal matrix elements (HUCKEL) or the rule of \citeit{Ammeter1978_JACS_3686} (MODHUCKEL) \label{enu:huckel}
\item the superposition of atomic potentials\cite{Lehtola2019_JCTC_1593} in (i) its original quadrature-based form, using pretabulated atomic contributions determined from fully numerical calculations with exchange-only LDA\cite{Lehtola2020_PRA_12516} (VSAP), as well as (ii) in its Gaussian fit form,\cite{Lehtola2020_JCP_144105} using the small (SAP S) or large (SAP L) GRASP fit of \citeitcomma{Lehtola2020_JCP_144105} or the Hartree--Fock (SAP HF) and Hartree--Fock--Slater potentials (SAP HFS) of \citeit{Laikov2020_TCA_17} \label{enu:sapfit}
\end{enumerate}
Since the SCF Fock matrix is in general spin-polarized, we spin-average it for open-shell systems for use in our algorithm: ${\bf F}_0 = ({\bf F}_0^\alpha + {\bf F}_0^\beta)/2$.
Next, the SAD and MINAO guesses only produce a density matrix; thereby their guess Fock matrix already includes a Fock build, and the resulting Fock matrix is again spin-averaged.
Similarly, the HUCKEL and MODHUCKEL guesses yield guess orbitals in an atomic minimal basis, and these guesses also require a Fock build and spin averaging to reach the full target basis.
In contrast, the various SAP guesses directly yield a spin-independent guess for the Fock matrix in the target basis, are thereby cheap to build, and were the original motivation for this work.

As some of the intermediate calculations in our exploratory calculations often landed on higher-lying extrema, which would have considerably complicated the analysis of our results, we decided to carry out the SCF benchmark for \cref{eq:descf} in the following manner.
First, the occupations for the irreducible symmetries are determined in the reference calculation with the full basis, and the occupations are then hardcoded for use in all dual-basis calculations.
Second, we also use robust convergence in the SCF, which consists of three conventional first-order SCF cycles with an applied level shift of 1.0 ${\rm E}_{\rm h}$, followed by a second-order solution with the co-iterative augmented Hessian (CIAH) method.\cite{Sun2017_CPL_291, Sun2017__}
We found that without the first-order minimization, the CIAH algorithm converges onto higher-lying solutions; similar issues in CIAH have also been recently documented by \citeitperiod{Greiner2026_JCTC_881}

\section{Results}
\label{sec:results}

Three quantities are relevant and interconnected in the analysis of the results.
First, $\Delta E^{\rm occ}$ of \cref{eq:deta} determines how quickly the occupied energy sum converges with the DB for a given initial guess ${\bf F}_0$.
Second, $\Delta E^{\rm SCF}$ in \cref{eq:descf} gauges the same for the variational SCF total energy, which is what we hope to be able to find a cheap approximation for.
Third, the size of the DB is what we hope to minimize; we study it as the fraction of kept basis functions (FKBF) for comparability across systems and basis sets of various sizes
\begin{equation}
  \label{eq:FKBF}
  {\rm FKBF}=N_{\rm{subbasis}}/N_{\text{full basis}}.
\end{equation}
Our analysis is based on these three quantities.

The automatic DB generation algorithm was run with a tolerance of $E^{\rm occ}_{\rm{tol}}=10^{-9}$ (see \cref{sec:implementation}).
We begin the analysis of the results by studying the connection between FKBF and $\Delta E^{\rm occ}$ to provide insight into how quickly the latter converges.
For the sake of conciseness we only present a representative set of figures that are essential for the discussion; the full set of figures is available in the supplemental information (SI).

The data shown in \cref{fig:c2h6-criterion} for the ethane molecule in the aug-pc-1 and aug-pc-3 basis sets demonstrates characteristic $({\rm FKBF},\Delta E^{\rm occ})$ plots, which exhibit fast initial convergence at small FKBF, followed by a slower asymptotic region extending to large FKBF, which then terminates in a sudden drop as the last important basis function is included in the DB.
The differences between the curves for the various initial guesses underline differences in the easiness to expand the distinct guesses' minimizing orbitals.

\begin{figure}
  \centering

  \begin{subfigure}{.75\linewidth}
    \centering
    \includegraphics[width=\linewidth]{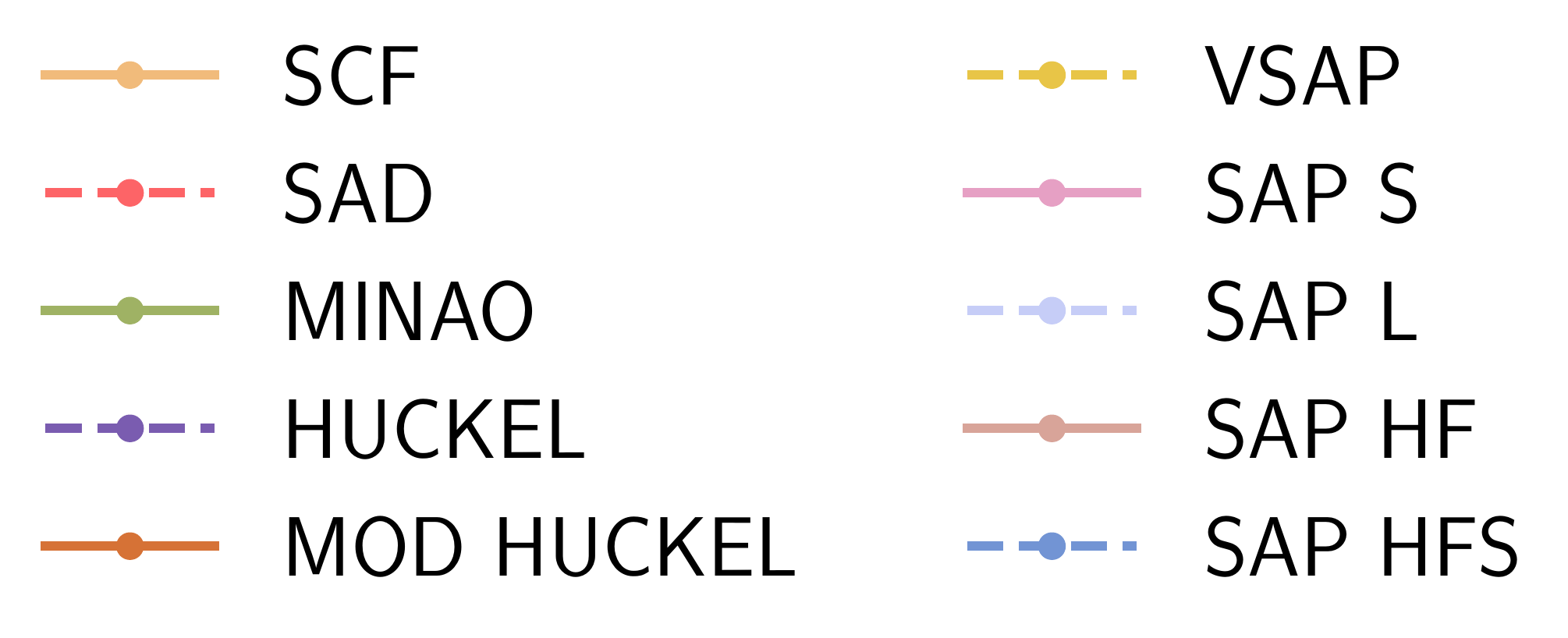}
  \end{subfigure}
  
  \begin{subfigure}{\linewidth}
    \centering
    \includegraphics[width=.98\linewidth]{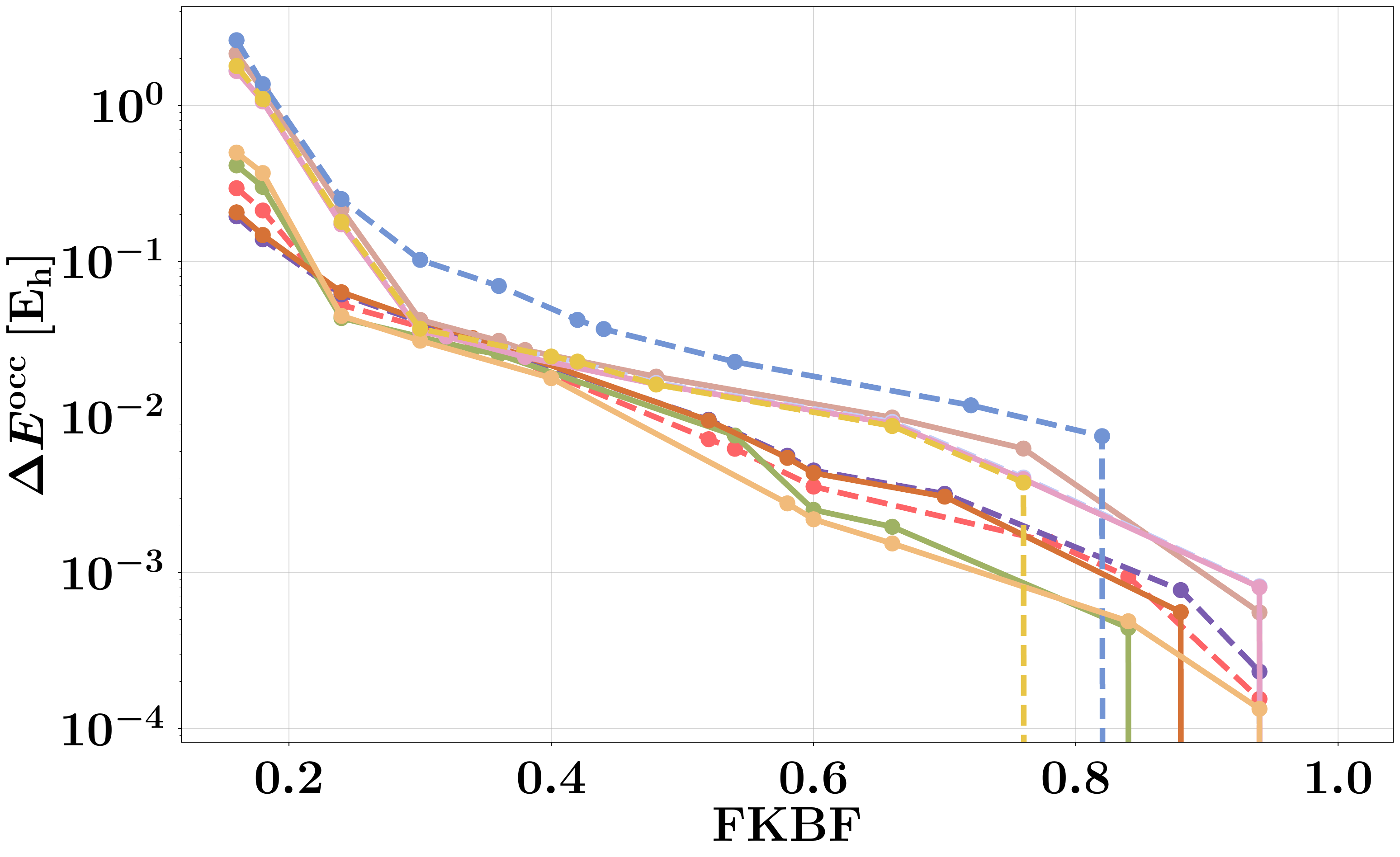}
    \subcaption{aug-pc-1}\label{fig:c2h6_criterion_apc1}
  \end{subfigure}
  
  \begin{subfigure}{\linewidth}
    \centering
    \includegraphics[width=.98\linewidth]{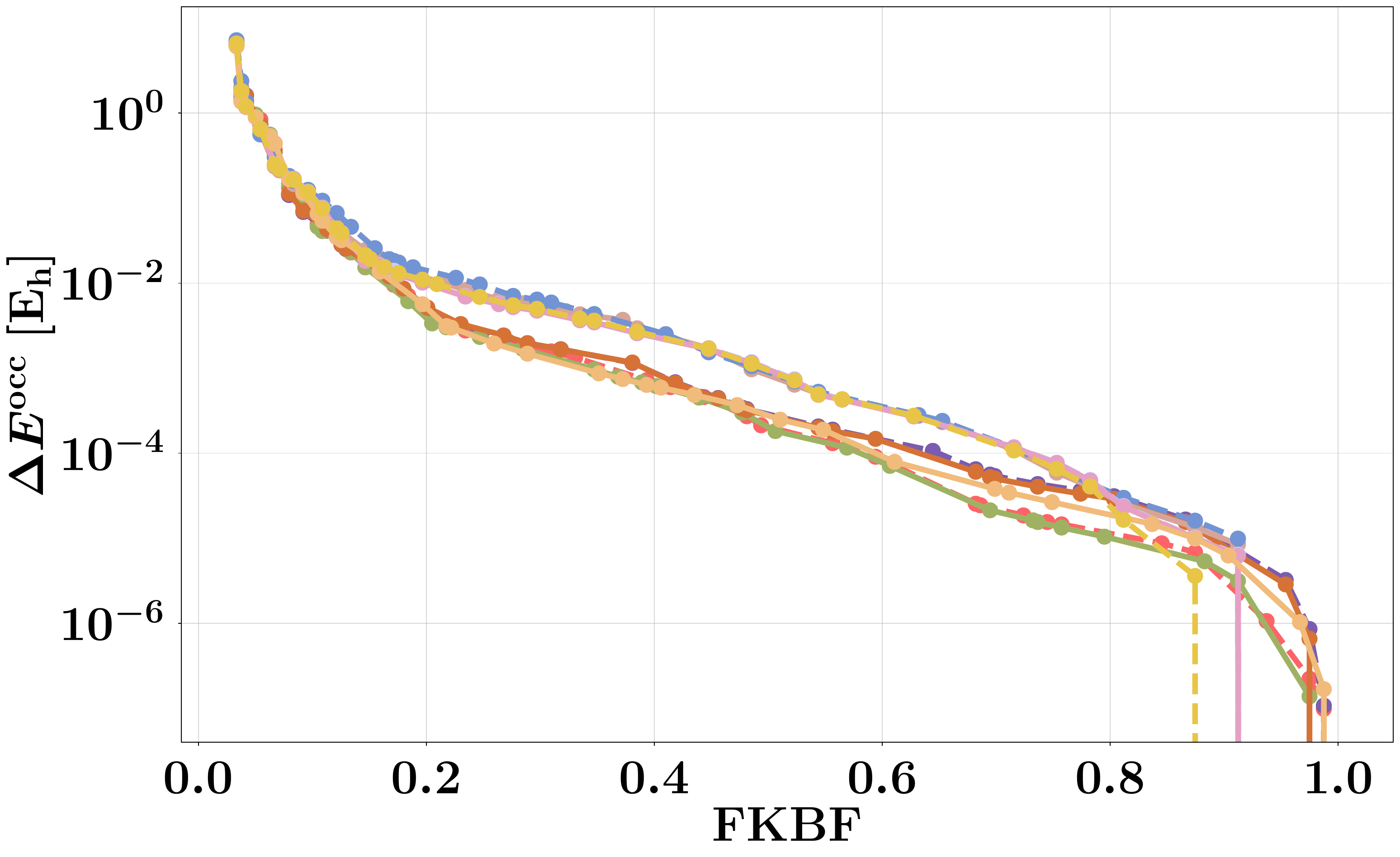}
    \subcaption{aug-pc-3}\label{fig:c2h6_criterion_apc3}
  \end{subfigure}

  \caption{
    The convergence of $E^{\rm occ}$ produced by the greedy algorithm for various initial guesses for the \ce{C2H6} molecule in the aug-pc-1 and aug-pc-3 basis sets.
  }
  \label{fig:c2h6-criterion}
\end{figure}

\begin{figure}
  \centering

  \begin{subfigure}{.75\linewidth}
    \centering
    \includegraphics[width=\linewidth]{figs/legend_2col.pdf}
  \end{subfigure}

  \begin{subfigure}{\linewidth}
    \centering
    \includegraphics[width=.98\linewidth]{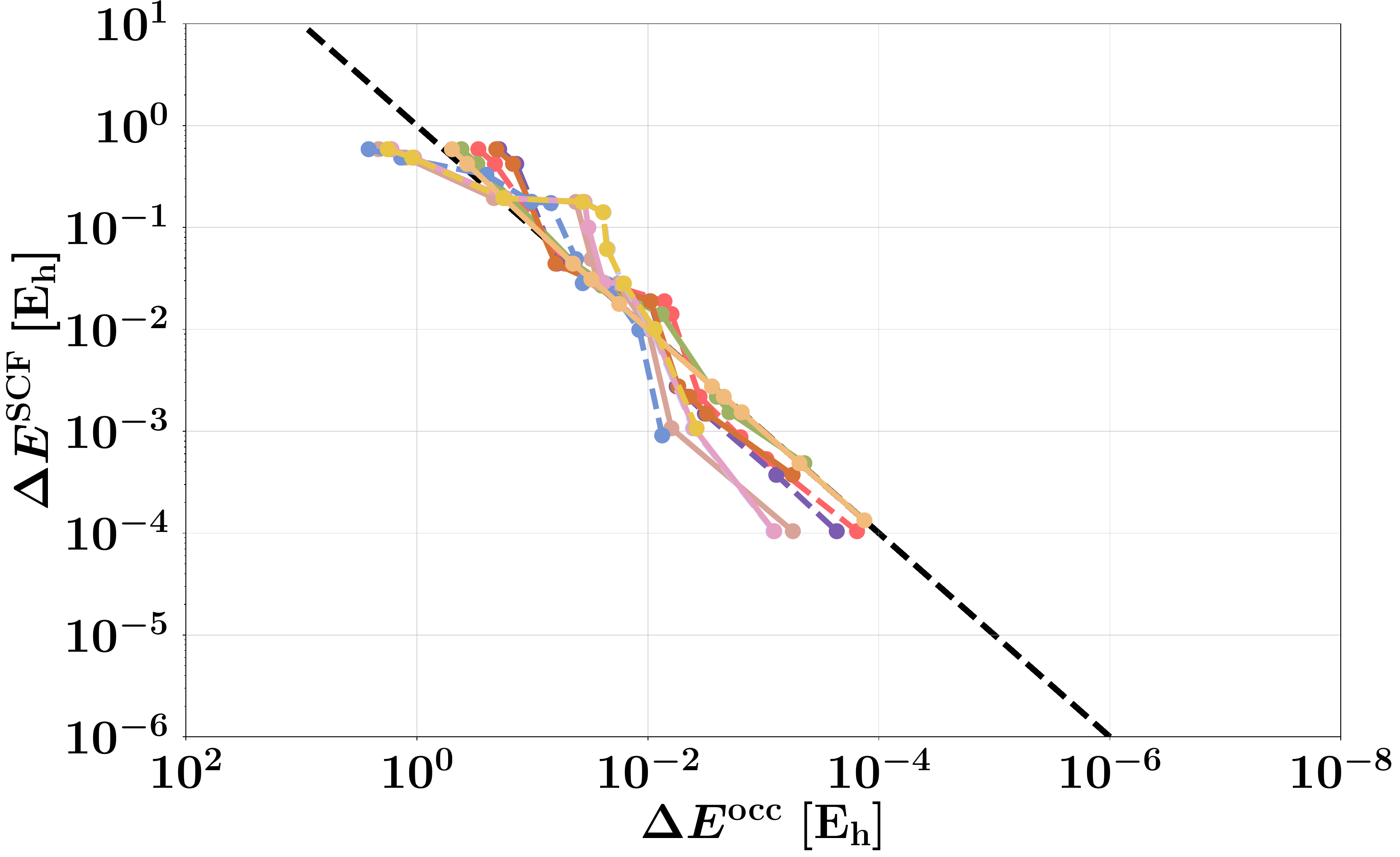}
    \subcaption{aug-pc-1}\label{fig:c2h6_crosscor_apc1}
  \end{subfigure}
  
  \begin{subfigure}{\linewidth}
    \centering
    \includegraphics[width=.98\linewidth]{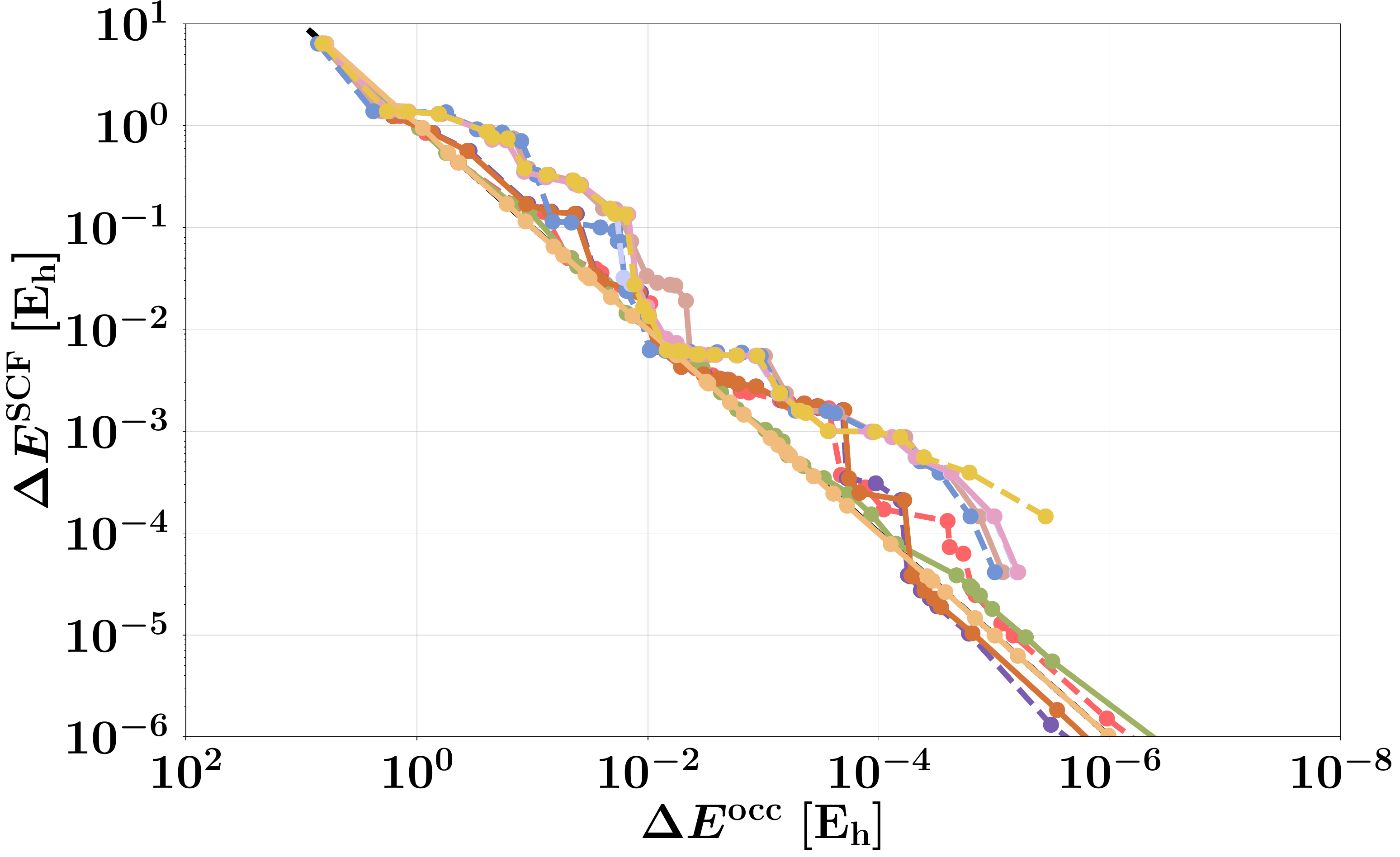}
    \subcaption{aug-pc-3}\label{fig:c2h6_crosscor_apc3}
  \end{subfigure}

  \caption{
    The relationship of $\Delta E^{\rm occ}$ and $\Delta E^{\rm SCF}$ for various initial guesses for the \ce{C2H6} molecule in the aug-pc-1 and aug-pc-3 basis sets.
    The dashed diagonal line represents the ideal one-to-one correspondence $\Delta E^{\rm SCF} = \Delta E^{\rm occ}$.
  }
  \label{fig:c2h6-crosscor}
\end{figure}

Now that we have an idea of how easily $\Delta E^{\rm occ}$ converges, we demonstrate its strong correlation with the variational SCF energy, verifying the goodness of $E^{\rm occ}$ for choosing the DB functions.
The analysis in \cref{fig:c2h6-crosscor} confirms that $\Delta E^{\rm occ}$ indeed works well as a proxy for $\Delta E^{\rm SCF}$, as the error in $\Delta E^{\rm occ}$ typically predicts the error in the SCF energy within an order of magnitude.
Although the differences between the curves in \cref{fig:c2h6-crosscor} mix two sources of error (the quality of the initial guess $\vb{F}_0$ and the quality of our greedy selection for a given $\vb{F}_0$), it is already noteworthy that the reference SCF curve is close to ideal and that the SAD and MINAO guesses are close to the ideal behavior.

\begin{figure*}
  \centering  
  \begin{subfigure}{\linewidth}
    \centering
    \includegraphics[width=.8\linewidth]{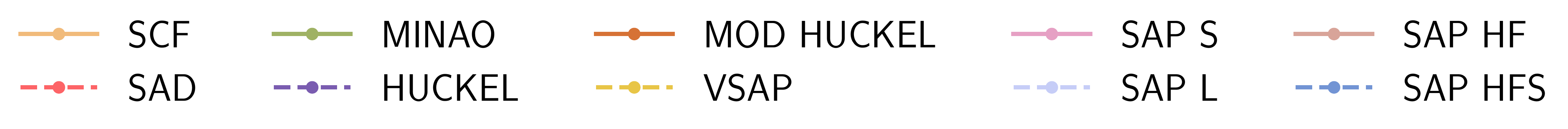}
    \includegraphics[width=.44\linewidth]{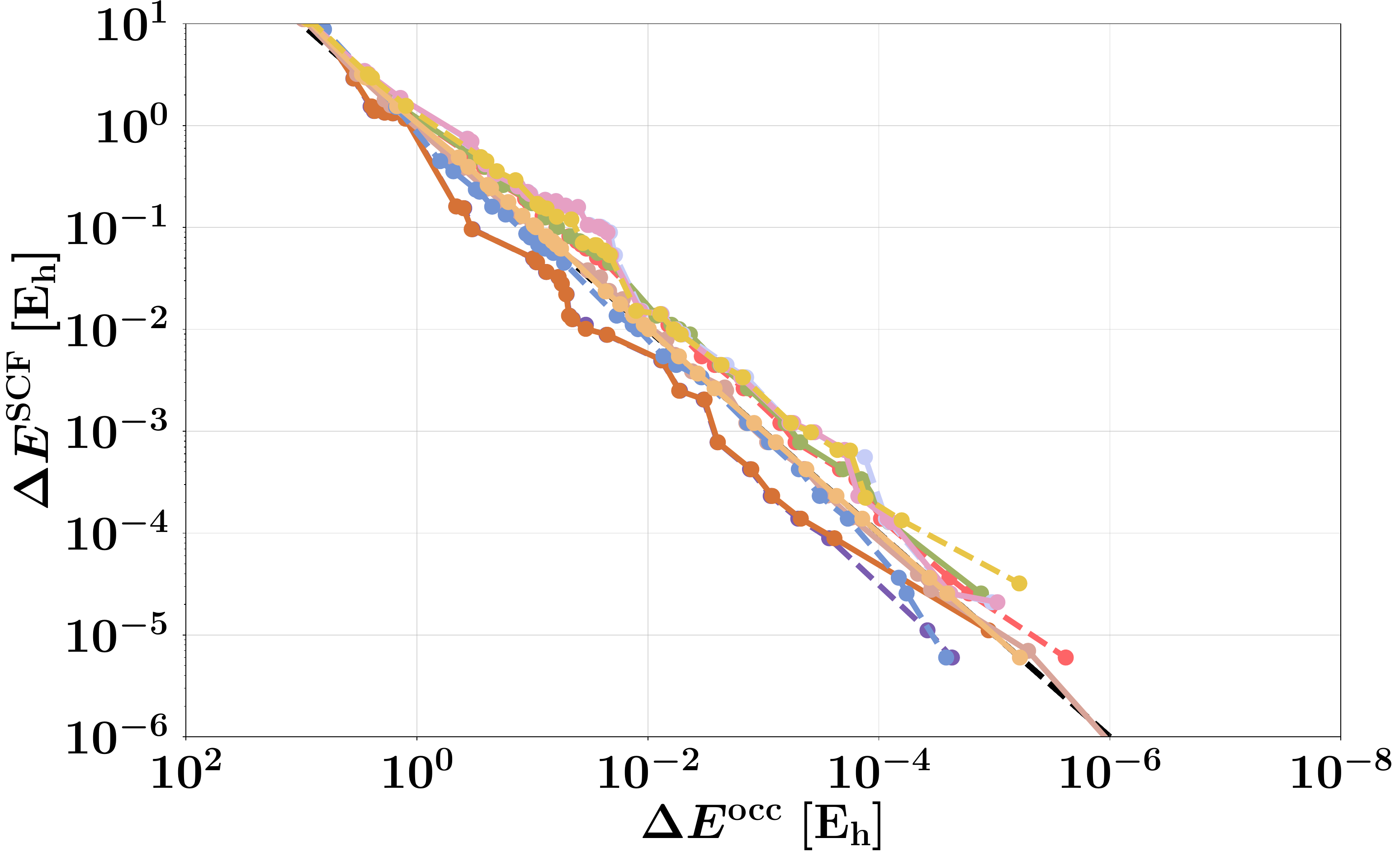}
    \hfill
    \includegraphics[width=.44\linewidth]{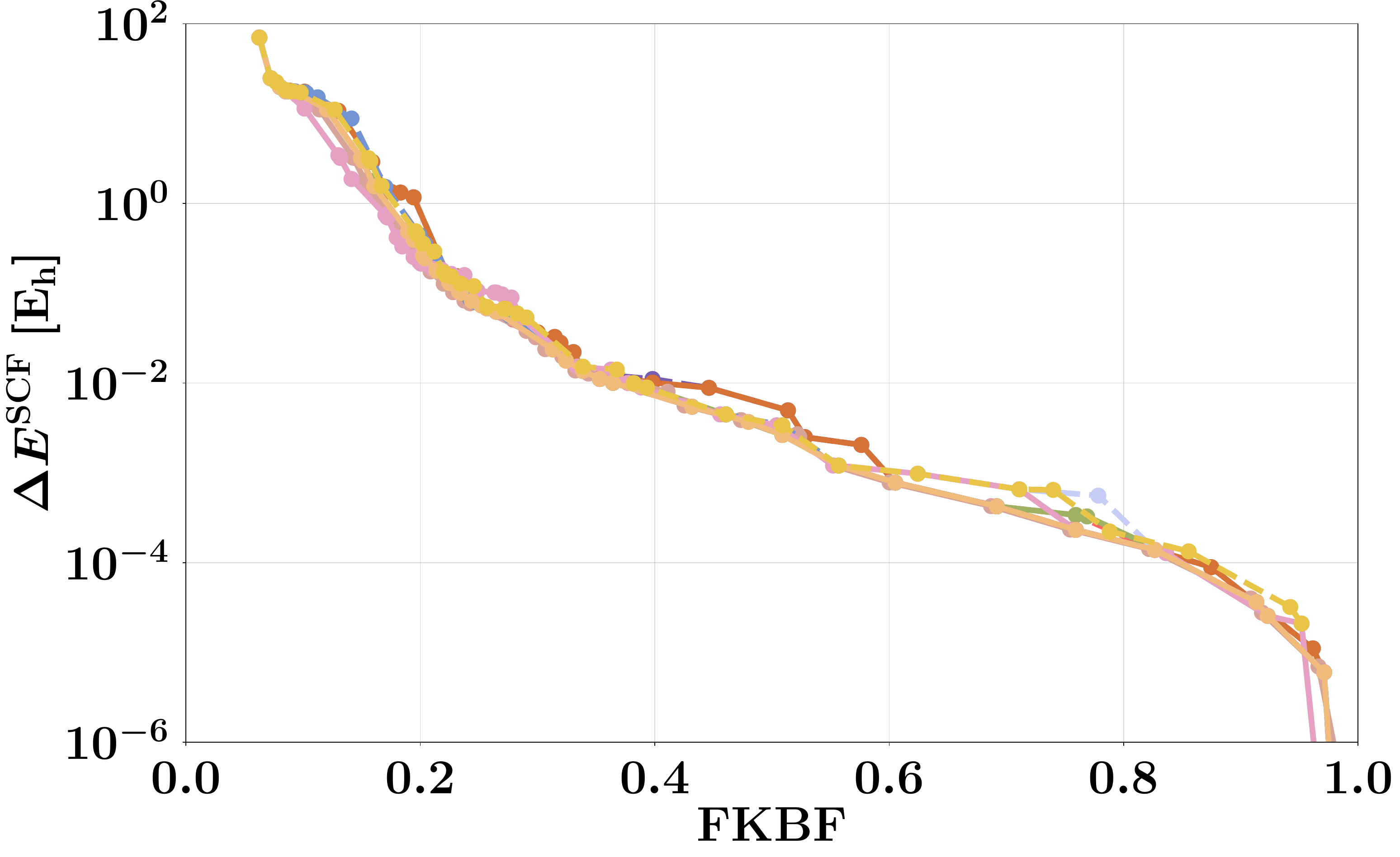}
    \subcaption{\ce{SF6}}
    \label{fig:sf6}
  \end{subfigure}
  \begin{subfigure}{\linewidth}
    \centering
    \includegraphics[width=.44\linewidth]{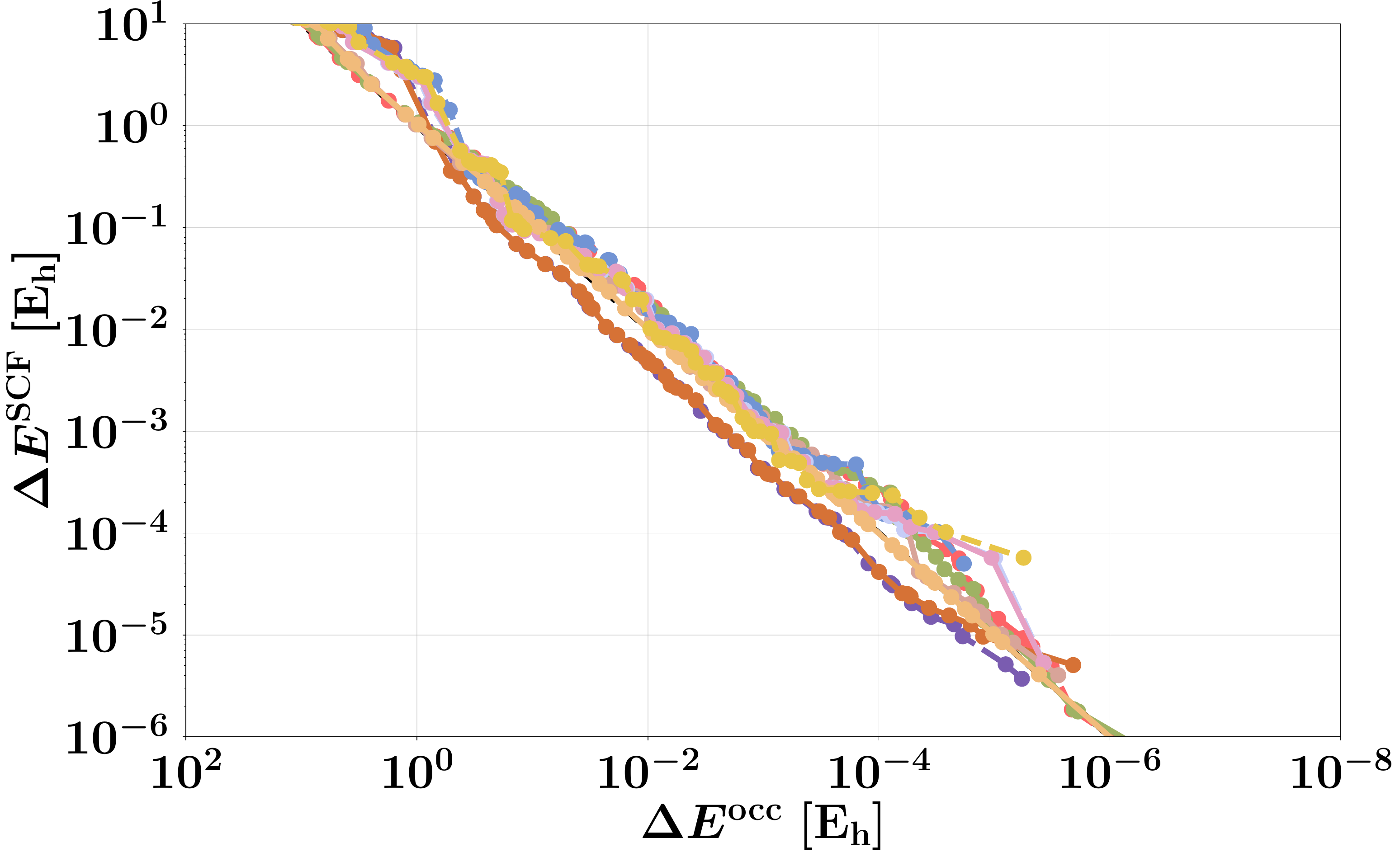}
    \hfill
    \includegraphics[width=.44\linewidth]{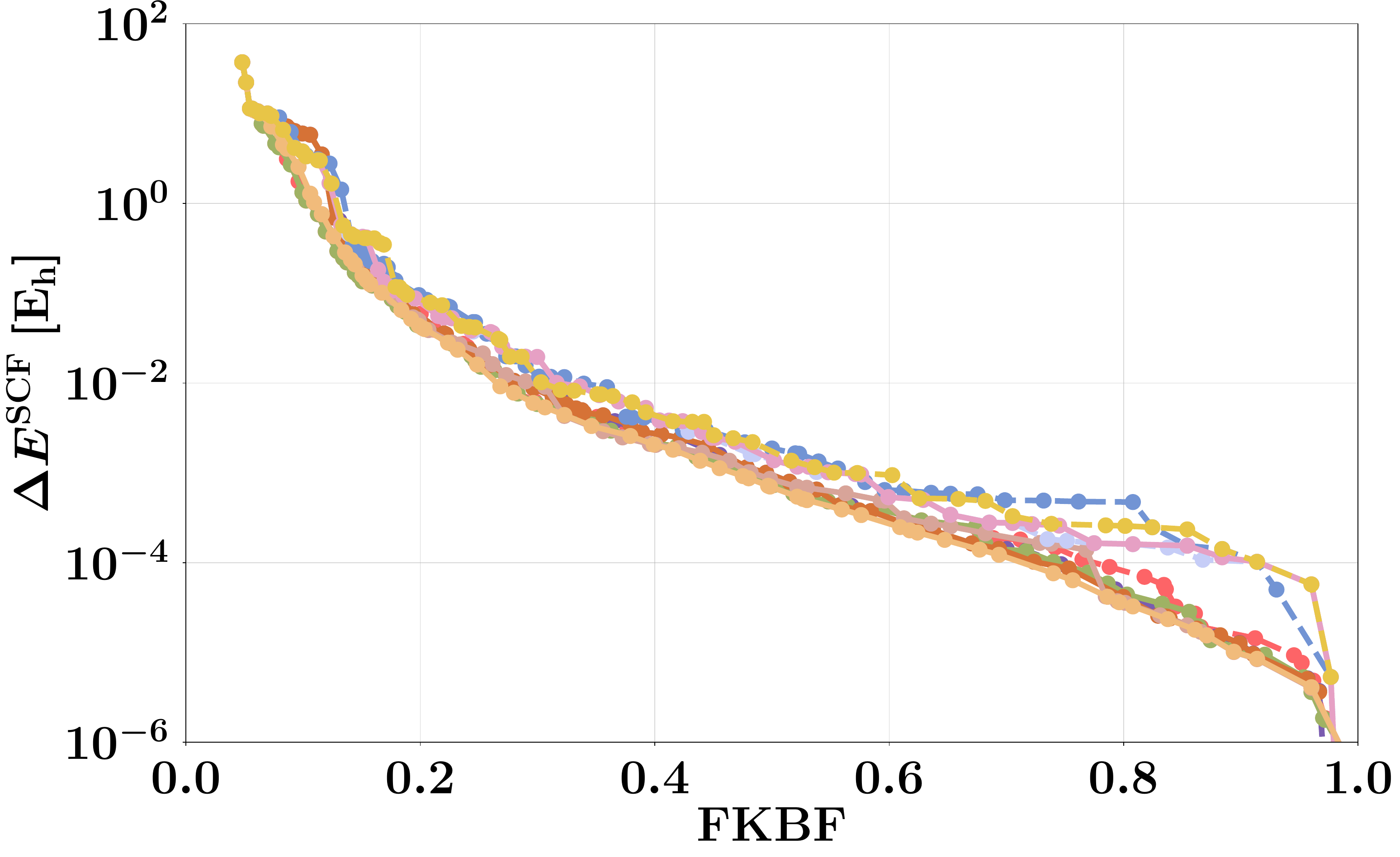}
    \subcaption{\ce{BeF2O2H4}}
    \label{fig:bef2o2h4}
  \end{subfigure}
  \begin{subfigure}{\linewidth}
    \centering
    \includegraphics[width=.44\linewidth]{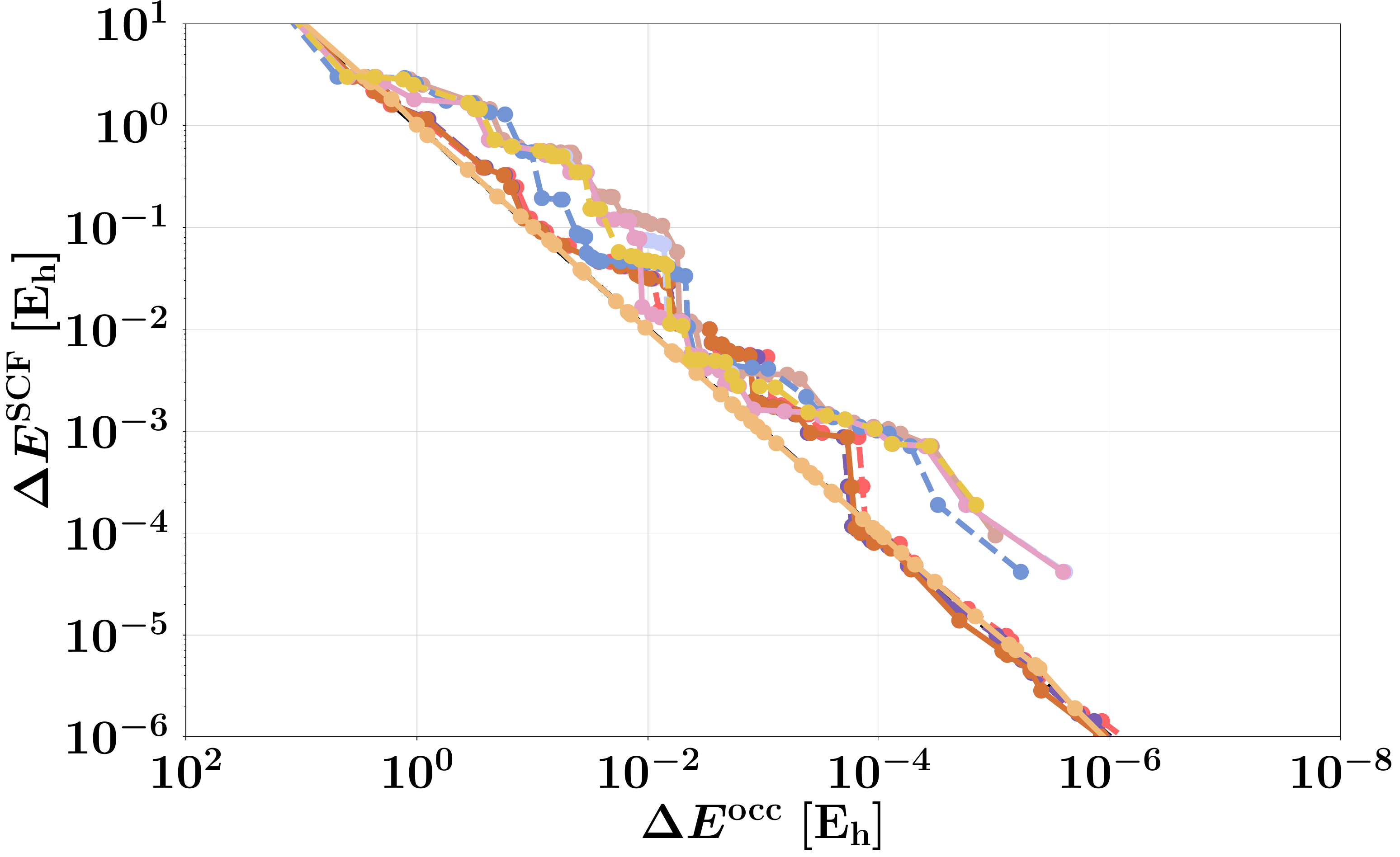}
    \hfill
    \includegraphics[width=.44\linewidth]{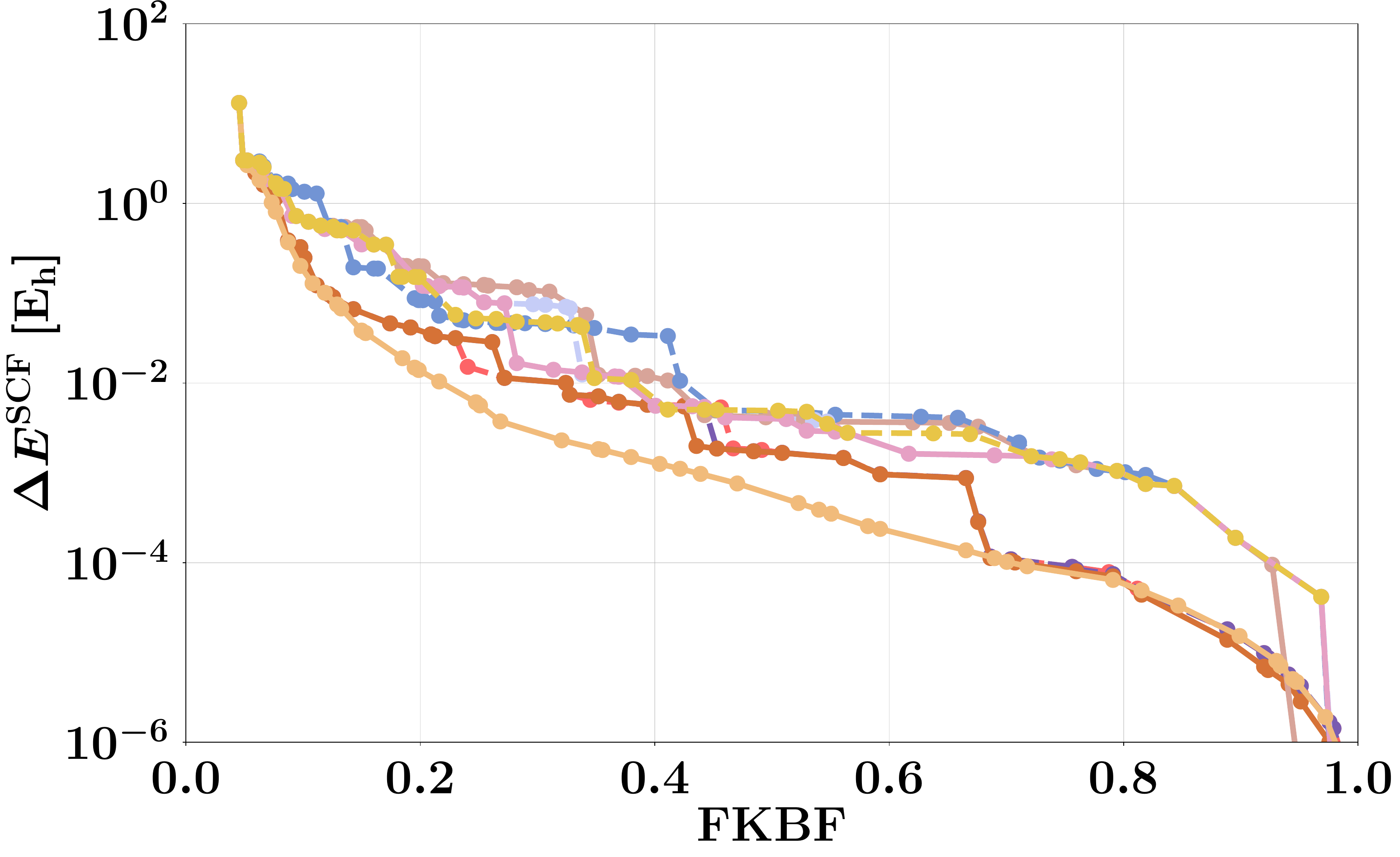}
    \subcaption{\ce{Li4C4H12}}
    \label{fig:li4c4h12}
  \end{subfigure}
  \begin{subfigure}{\linewidth}
    \centering
    \includegraphics[width=.44\linewidth]{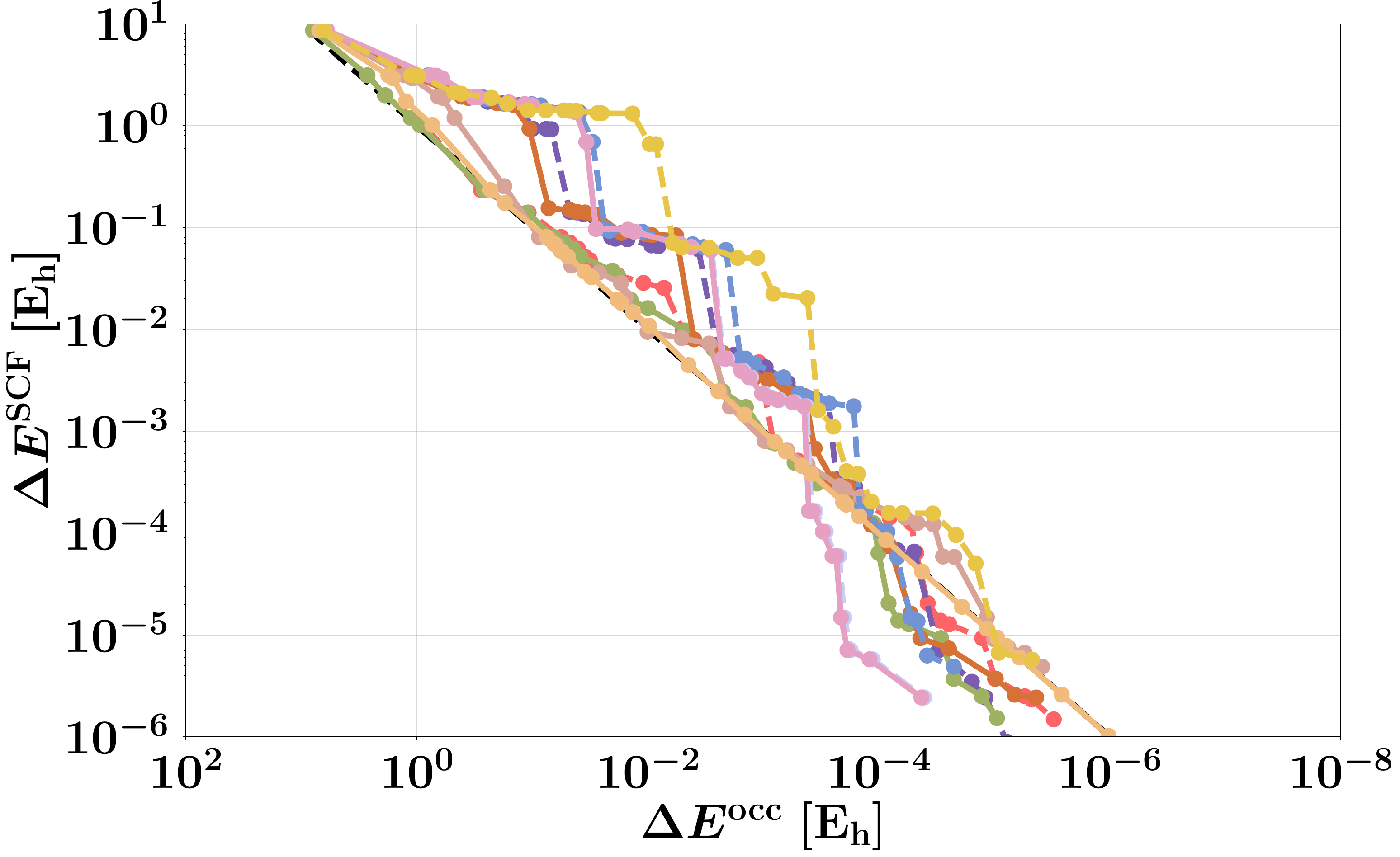}
    \hfill
    \includegraphics[width=.44\linewidth]{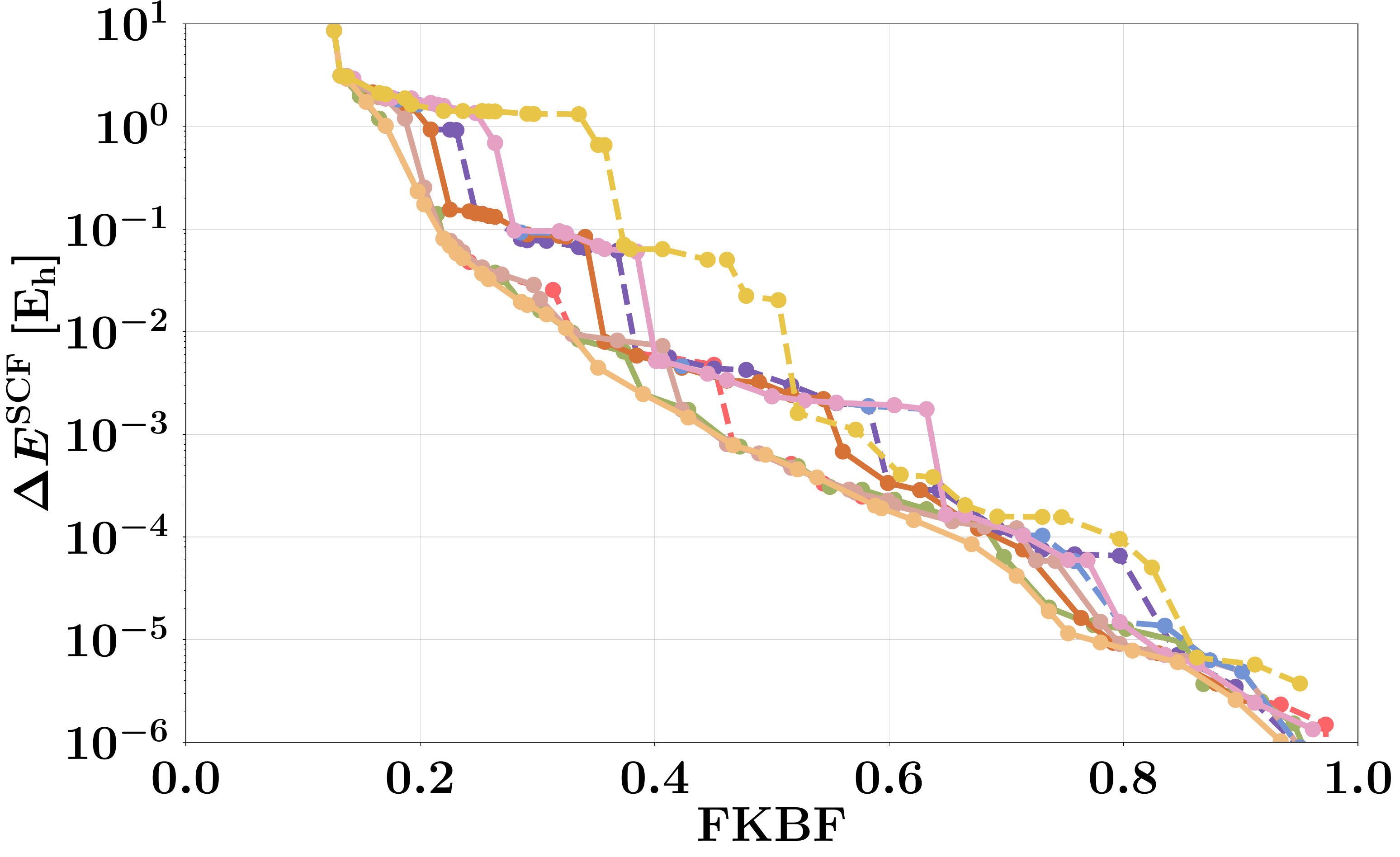}
    \subcaption{\ce{GaO}}\label{fig:gao}
  \end{subfigure}
  \caption{
    Truncation errors in the converged DB set energies to the full basis energy as a function of the error in the occupied orbital energies (left panel) and the fraction of kept basis functions (right panel) for \ce{SF6} (\cref{fig:sf6}), \ce{BeF2O2H4} (\cref{fig:bef2o2h4}), \ce{Li4C4H12} (\cref{fig:li4c4h12}), and \ce{GaO} (\cref{fig:gao}) in the aug-pc-3 basis set.
    Note the inverted $x$-axis for the figures in the left panel.
    The dashed diagonal line indicates the optimal one-to-one correspondence line.
  }\label{fig:comp_fig1}
\end{figure*}

\begin{figure*}
  \centering

  \begin{subfigure}{\linewidth}
    \centering
    \includegraphics[width=.8\linewidth]{figs/legend.pdf}
    \includegraphics[width=.44\linewidth]{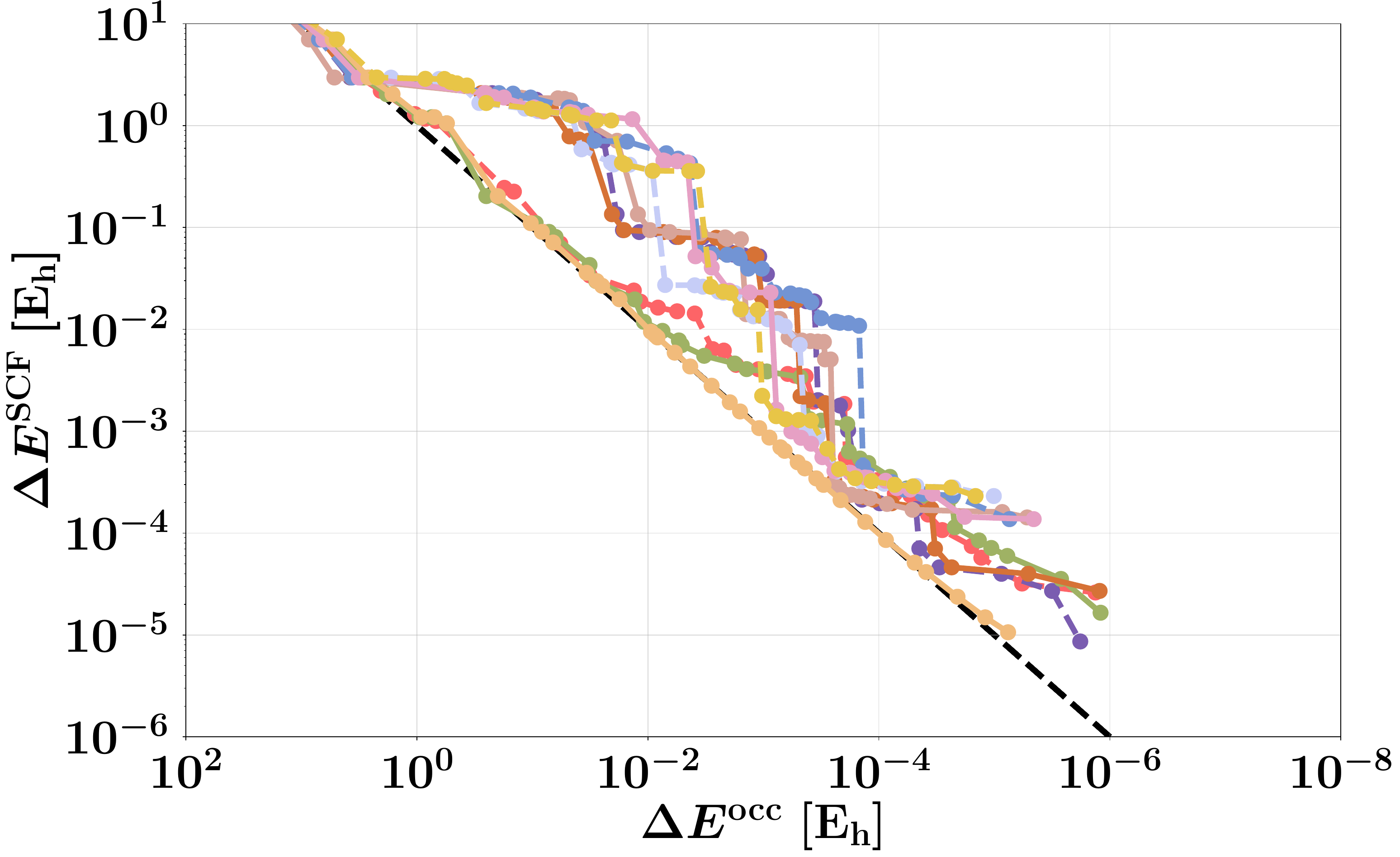}
    \hfill
    \includegraphics[width=.44\linewidth]{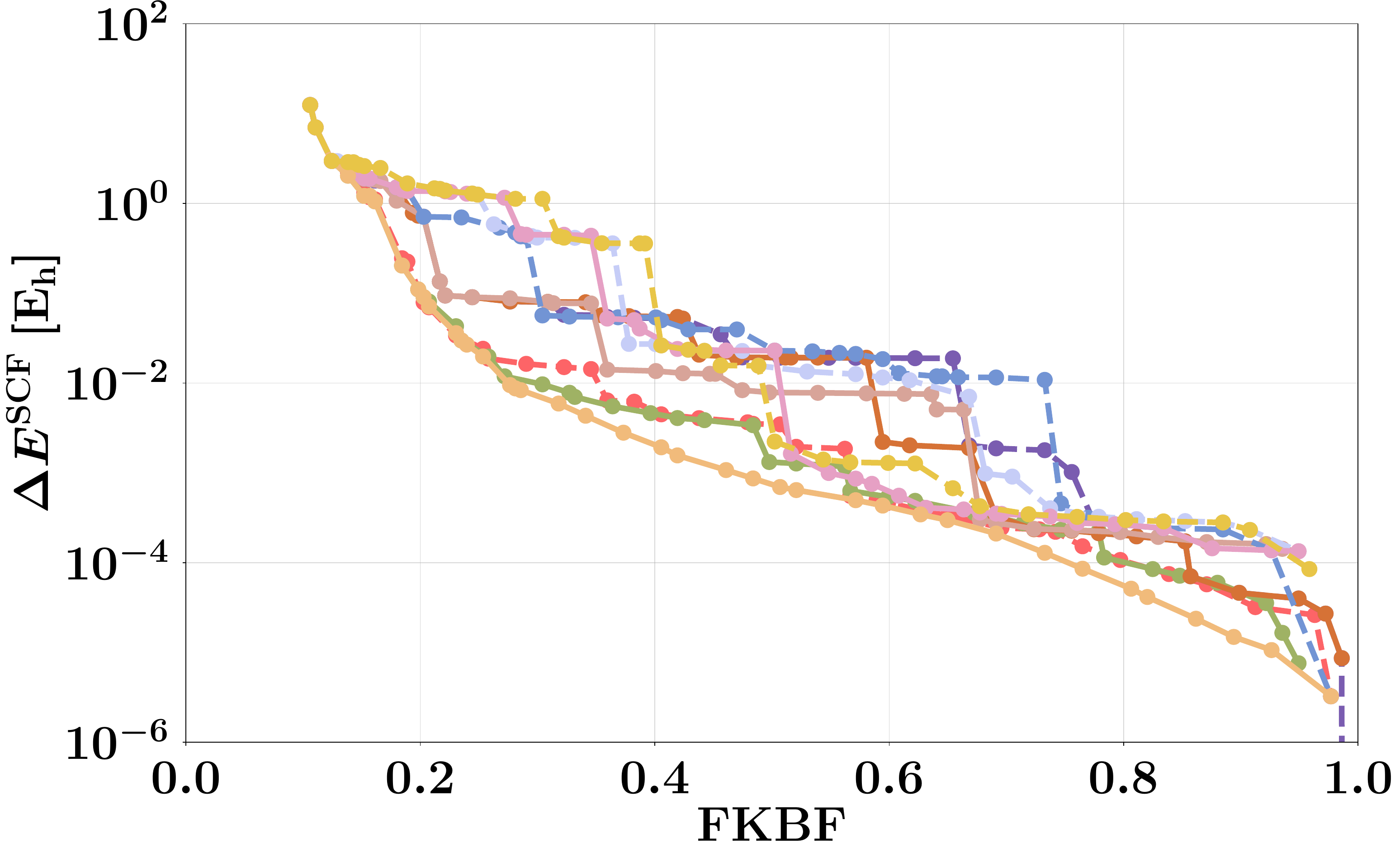}
    \subcaption{\ce{TiO}}
    \label{fig:tio}
  \end{subfigure}
  \begin{subfigure}{\linewidth}
    \centering
    \includegraphics[width=.44\linewidth]{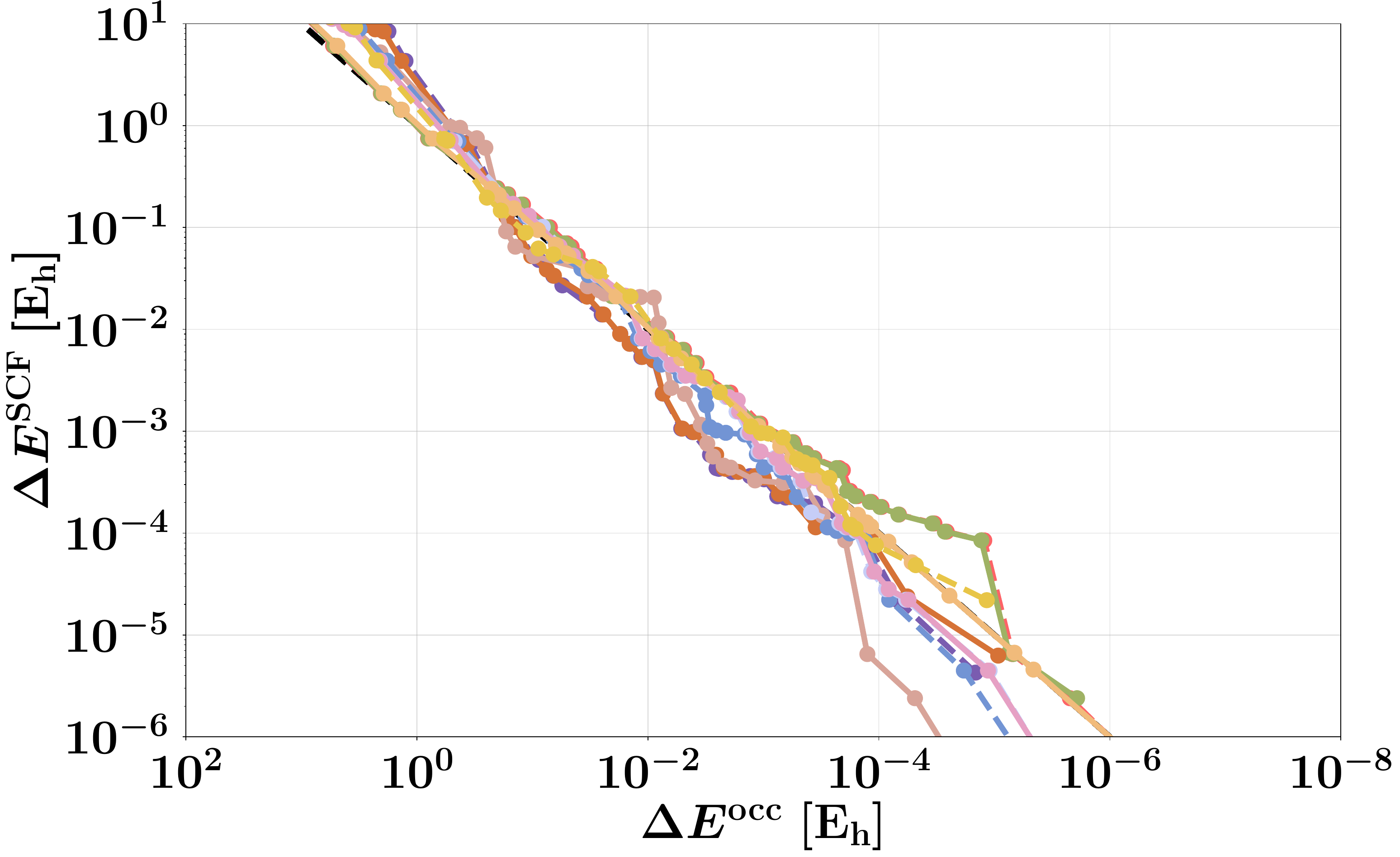}
    \hfill
    \includegraphics[width=.44\linewidth]{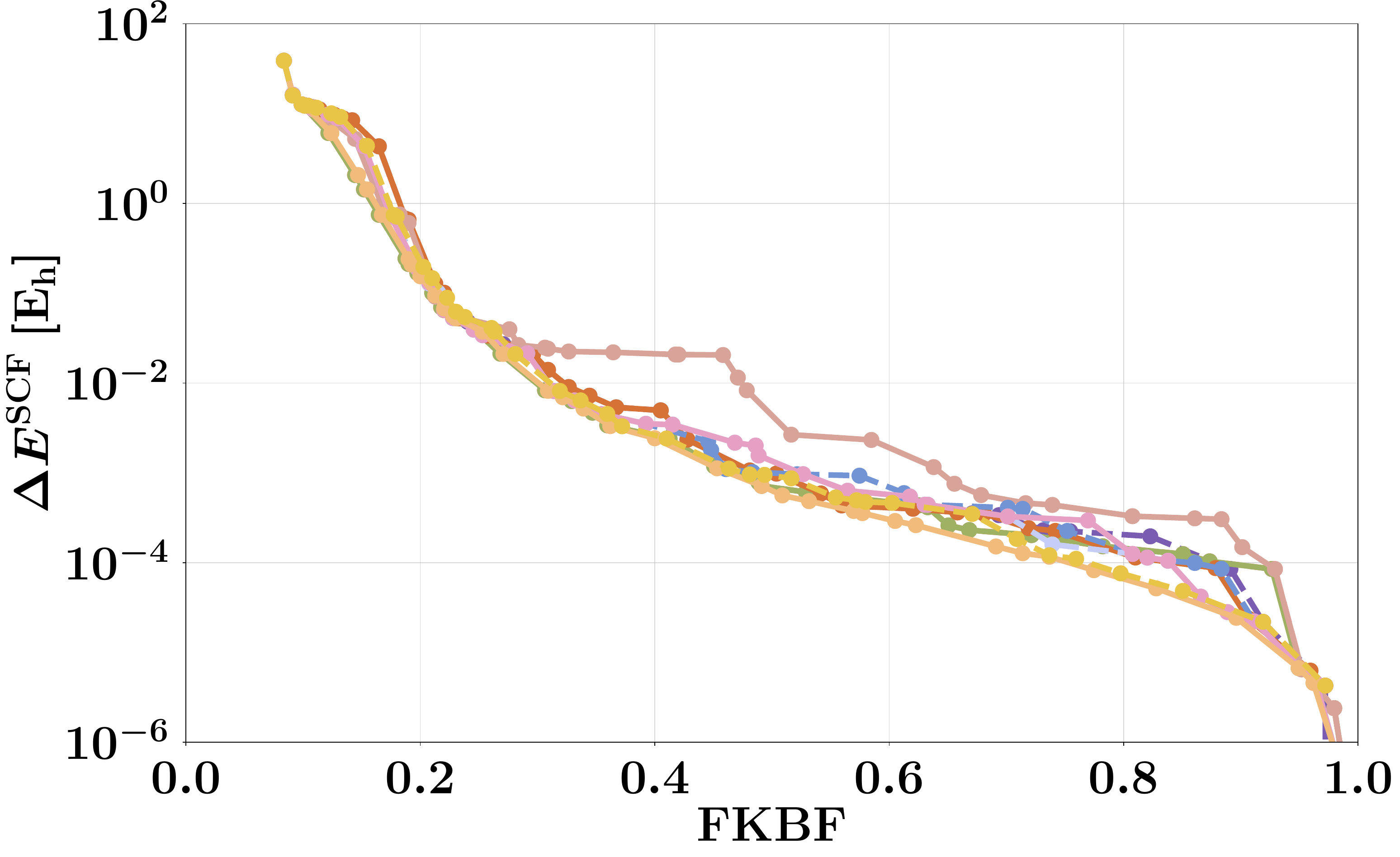}
    \subcaption{\ce{CrF3}}
    \label{fig:crf3}
  \end{subfigure}
  \begin{subfigure}{\linewidth}
    \centering
    \includegraphics[width=.44\linewidth]{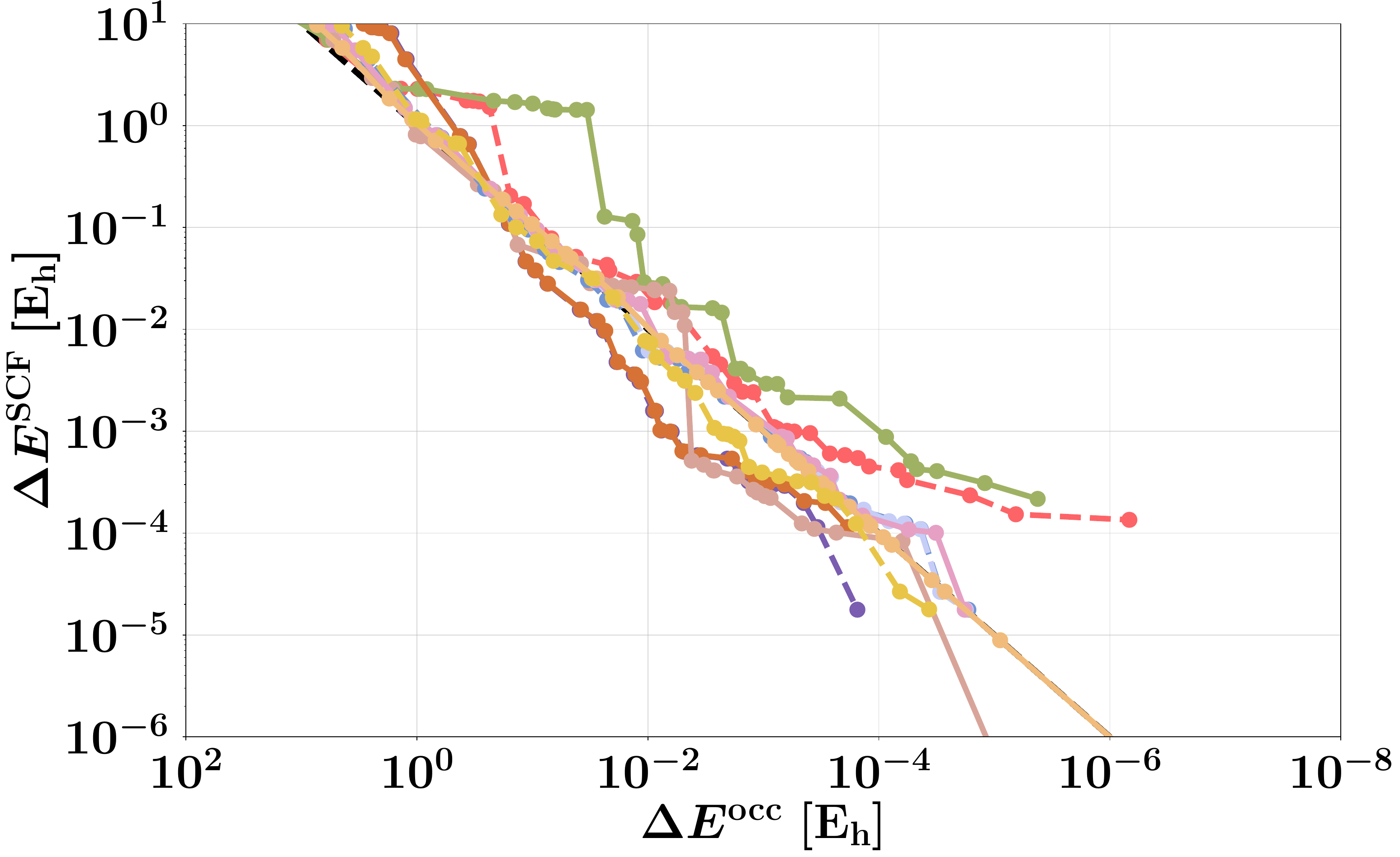}
    \hfill
    \includegraphics[width=.44\linewidth]{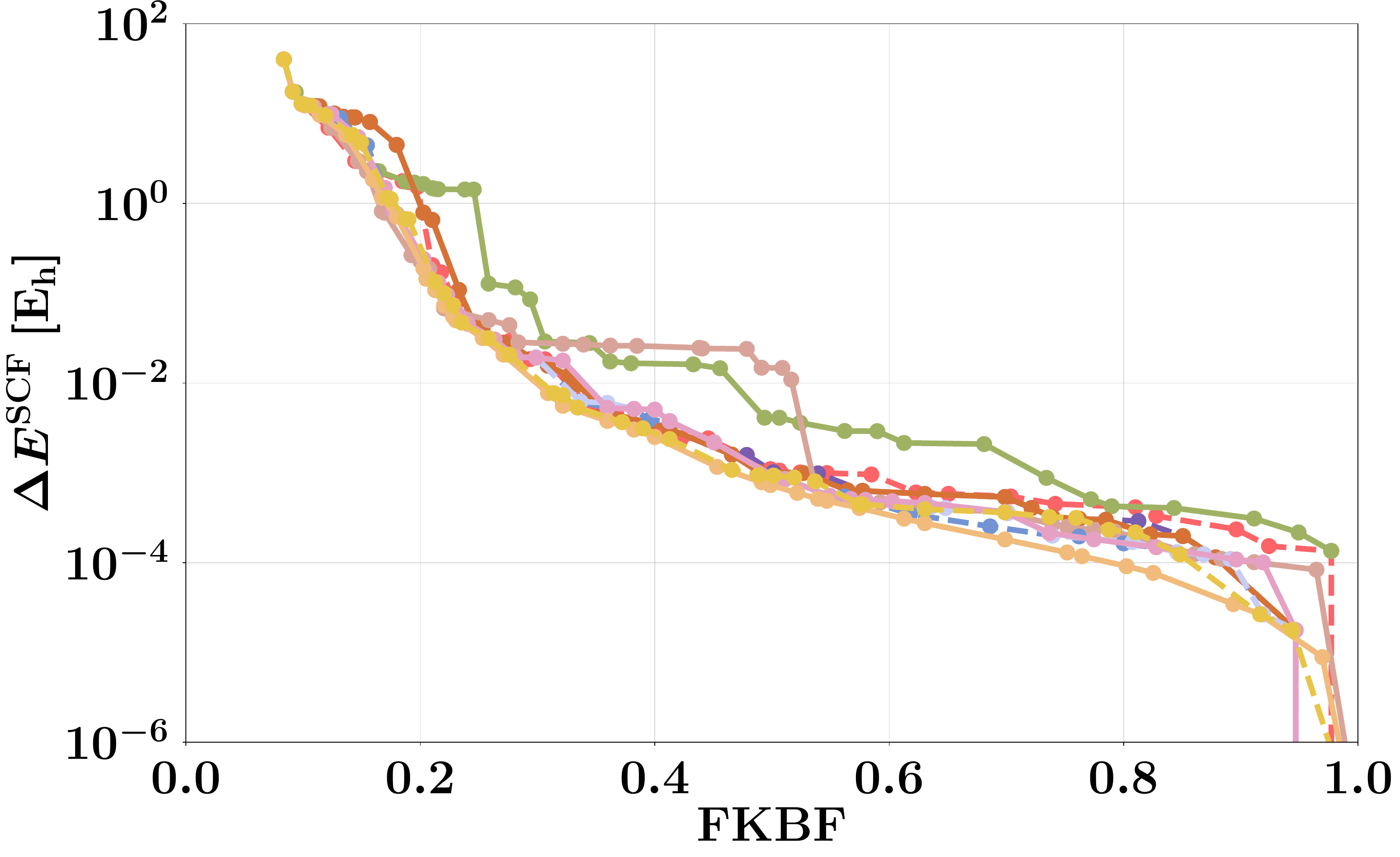}
    \subcaption{\ce{CoF3}}
    \label{fig:cof3}
  \end{subfigure}
  \begin{subfigure}{\linewidth}
    \centering
    \includegraphics[width=.44\linewidth]{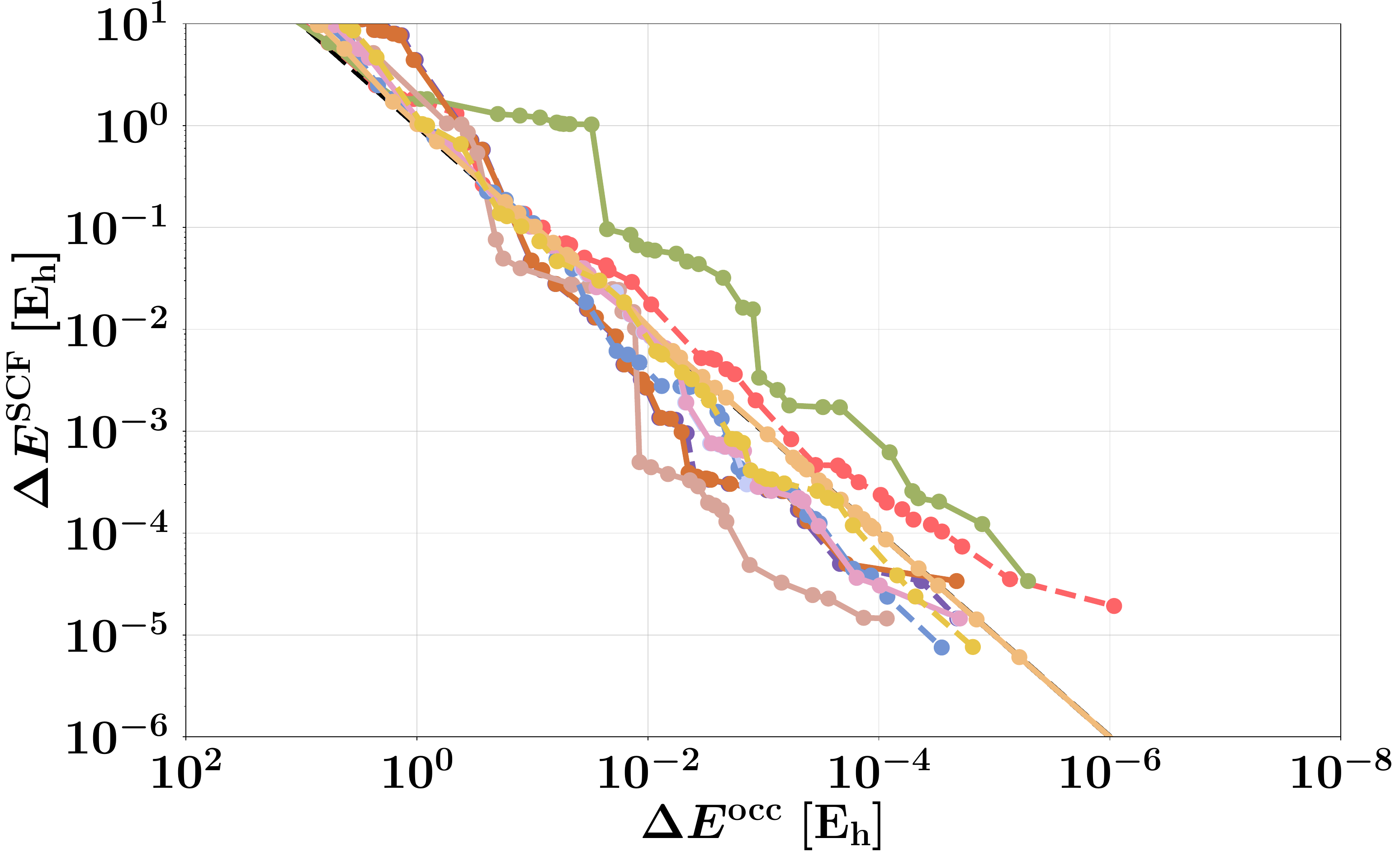}
    \hfill
    \includegraphics[width=.44\linewidth]{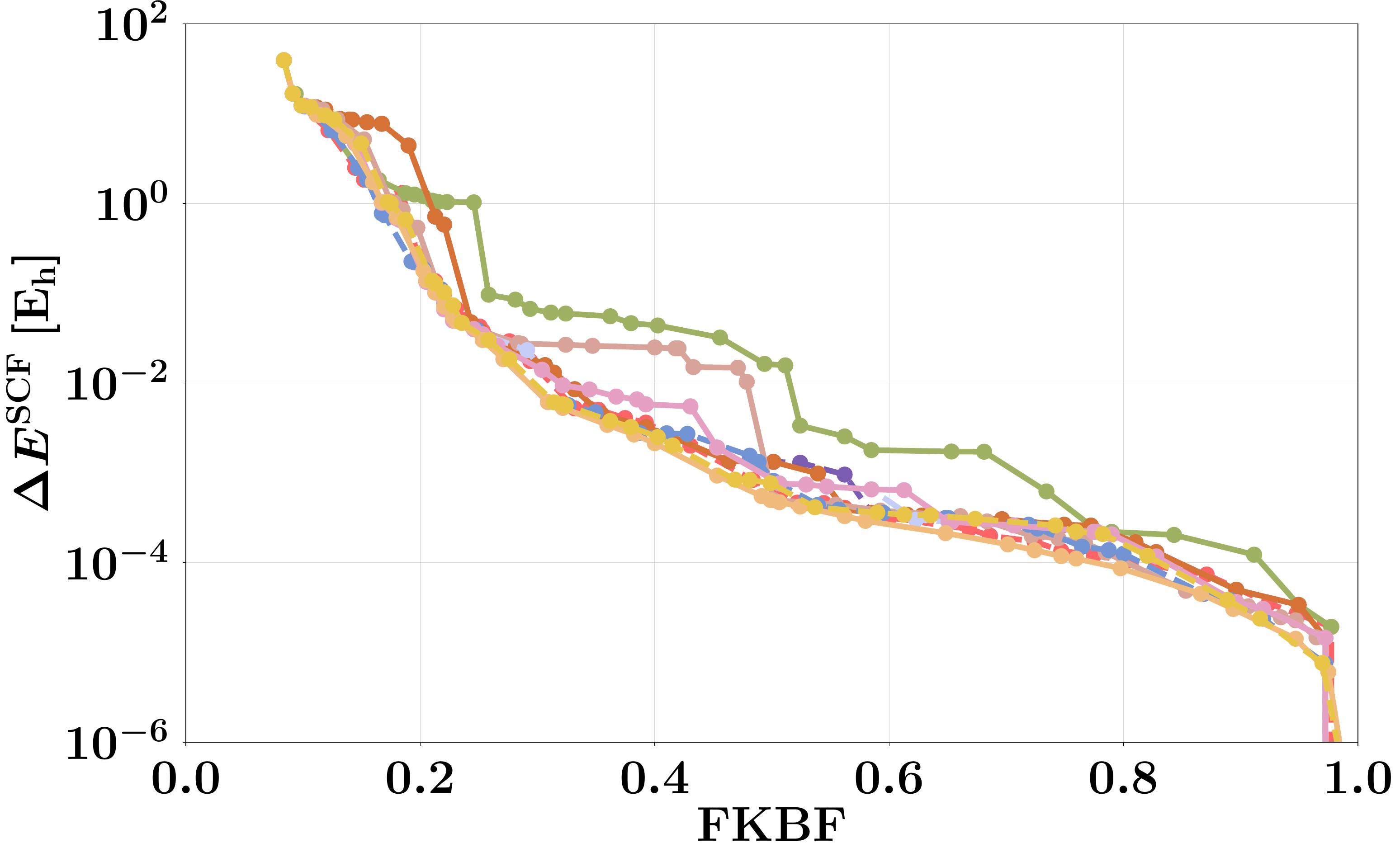}
    \subcaption{\ce{FeF3}}
    \label{fig:fef3}
  \end{subfigure}

  \caption{
    Truncation errors in the converged DB set energies to the full basis energy as a function of the error in the occupied orbital energies (left panel) and the fraction of kept basis functions (right panel) for \ce{TiO} (\cref{fig:tio}), \ce{CrF3} (\cref{fig:crf3}), \ce{CoF3} (\cref{fig:cof3}), and \ce{FeF3} (\cref{fig:fef3}) in the aug-pc-3 basis set.
    Note the inverted $x$-axis for the figures in the left panel.
    The dashed diagonal line indicates the optimal one-to-one correspondence line.
  }\label{fig:comp_fig2}
\end{figure*}

At this stage, we extend the analysis to the \ce{SF6}, \ce{BeF2O2H4}, \ce{Li4C4H12}, and \ce{GaO} molecules shown in \cref{fig:comp_fig1} (left panels), as well as to the \ce{TiO}, \ce{CrF3}, \ce{CoF3} and \ce{FeF3} molecules shown in \cref{fig:comp_fig2} (left panels) to exemplify variation in convergence profiles as well as to cover various spin states.
The first three systems have a singlet ground state and span various sizes and light and heavy atoms.
\ce{SF6} is chosen specifically for its notoriously slowly converging atomization energy with respect to the size of the basis set.\cite{Kendall1992_JCP_6796,Weigend2005_PCCP_3297,Lehtola2020_JCP_134108,Schwalbe2020_JCP_84104}
The remaining systems have ground states ranging from doublet to sextet.
The overall trend across all systems is the same as that already observed for $E^\text{occ}$.
The truncation error of the SCF energy decreases rapidly in the initial phase.
Then, a plateau region is encountered, where the error decreases much more slowly, and in the final phase the error collapses when the last important function is included.

Smooth convergence is observed when the converged Fock matrix for the target basis is used to pick the DB.
However, when an initial guess is used to choose the DB, the convergence is not always smooth: the greedy search sometimes results in a staggered profile in the truncation error, visible in e.g. \cref{fig:bef2o2h4,fig:li4c4h12,fig:gao,fig:tio,fig:crf3,fig:cof3,fig:fef3}.
The analysis of these results is supplemented by plots showing the relation between the FKBF and $\Delta E^{\rm SCF}$ in the right panel of \cref{fig:comp_fig1,fig:comp_fig2}, showing how the SCF energy converges with respect to the size of the DB.
Since both FKBF and $\Delta E^{\rm occ}$ are monotone, the double view of $\Delta E^{\rm SCF}$ as a function of the FKBF as well as $\Delta E^{\rm occ}$ allows discerning better the changes in the size of the DB even when $\Delta E^{\rm occ}$ is stagnated.

Our working theory was that the sharp drops were due to additions of higher polarization functions.
Instead, our analysis identified that the sharp drops were usually due to the addition of diffuse $p$ functions with exponents ranging from around 0.1 to around 1.0.
For example, inspection of the calculations on \ce{CrF3} (\cref{fig:crf3}), \ce{CoF3} (\cref{fig:cof3}) and \ce{FeF3} (\cref{fig:fef3}) with the SAP HF potential reveals that no $p$-functions on any of the atoms are selected within the plateau regions lying higher compared to the other trajectories.
The addition of a $p$-function on the ligand atoms then dramatically decreases the truncation error.
Also $d$- and $f$-functions occasionally, although more rarely, led to similar yet less pronounced sharp drops, such as the one visible for the \ce{CrF3} SAP HF curve in \cref{fig:crf3} (right panel).
Since the greedy algorithm exhibits near-ideal behavior with the reference SCF Fock matrix ``guess'', these findings are emblematic of shortcomings in the initial guesses: a guess that is too far from correct is unable to predict the energetic importance of the various basis functions.

\section{Discussion and Summary}
\label{sec:discussion}

We have described an algorithm that automatically generates a system-adapted dual basis (DB) set for any orbital basis set at any desired precision, whose motivation was the efficient determination of DB approaches to electronic structure calculations.
The algorithm's only inputs are the overlap matrix and a guess Fock matrix ${\bf F}_0$ that is used to rank the basis functions by energetic importance.
Seeded with a minimal basis, the algorithm greedily adds functions that lower the sum of the occupied orbitals' energies $\Delta E^{\rm occ}$ (which is a monotonic function) until the predefined convergence threshold is met.

We proved and demonstrated that $\Delta E^{\rm occ}$ converges smoothly, and that it is a reliable proxy for the error in the SCF energy $\Delta E^{\rm SCF}$, the two being close in magnitude when the self-consistent Fock matrix in the full basis is used to drive the selection of the DB.
However, when combined with an approximate initial guess instead of the exact Fock matrix, the basis selection can plateau, highlighting deficiencies in the currently available initial guesses.
Moreover, the achieved compactification of the basis turned out surprisingly slow: even though our expectation was that since the variational theorem would have more freedom to work in a large basis, it might lead to fast convergence in large basis sets, the region of slow asymptotic convergence is wide.
This can be tentatively interpreted as the inherent slow convergence of atomic-orbital expansions of electronic structure.

The above observations also extend beyond our study.
\citeit{Lara2026_JCTC_3327} have recently reported a lossy compression approach for atomic orbital basis sets, based on the eigendecomposition of the atomic diagonal blocks of the density matrix.
Their scheme, so far only implemented as a proof-of-concept, will likewise hinge on the availability of an accurate initial guess.

\begin{acknowledgments}
We thank the Academy of Finland for financial support under project numbers 350282 and 353749, as well as CSC--IT Center for Science Ltd (Espoo, Finland) for computational resources.
\end{acknowledgments}

\section*{Data Availability}
The data that supports the findings of this study are available within the article and its supplementary material.

\section*{Conflict of Interest}
The authors have no conflicts to disclose.

\section*{Supplemental Information}
The full sets of (FKBF, $\Delta E^{\rm occ}$), (FKBF, $\Delta E^{\rm SCF}$) and $(\Delta E^{\rm occ},\Delta E^{\rm SCF})$ plots for the 185 molecules studied here with the aug-pc-$n$ basis sets with $n\in\{1,2,3\}$.
Each plot shows results with all 9 initial guesses and the SCF Fock matrix.

\bibliography{susi}

\end{document}